\documentclass[12pt]{article}
\pdfoutput=1
\usepackage[colorlinks=true,urlcolor=blue,anchorcolor=blue,citecolor=blue,filecolor=blue,linkcolor=blue,menucolor=blue,linktocpage=true,pdfproducer=medialab]{hyperref}
\usepackage{graphicx}
\usepackage{color,amsmath,amssymb,exscale,psfrag,epsfig}
\usepackage{cite,color,url}
\usepackage{pdfpages}
\usepackage{multirow}
\input epsf
\usepackage[utf8]{inputenc}

\newcommand{\m}[1]{\marginpar{{\tiny *}} }

\def\bea{\begin{eqnarray}}
\def\eea{\end{eqnarray}}

\newcommand\f[2]{\frac{#1}{#2}}

\catcode`\@=11
\def\lsim{\mathrel{\mathpalette\@versim<}}
\def\gsim{\mathrel{\mathpalette\@versim>}}
\def\@versim#1#2{\vcenter{\offinterlineskip
\ialign{$\m@th#1\hfil##\hfil$\crcr#2\crcr\sim\crcr } }}
\catcode`\@=12
\catcode`@=12
\begin{document}
\thispagestyle{empty}
\begin{flushright}
ICAS 026/17 \\
\end{flushright}
\vspace{1cm}
\begin{center}
{\Large \bf ttbb as a probe of New Physics at the LHC} \\
\vspace{0.6in}
{\bf Ezequiel \'Alvarez and Mariel Est\'evez}
\vspace{0.2in} \\
{\sl International Center for Advanced Studies (ICAS), ECyT, UNSAM, Campus Miguelete \\
25 de Mayo y Francia, (1650) Buenos Aires, Argentina
}
\end{center}
\vspace{1in}

\begin{abstract}
We study the $t\bar t b \bar b$ final state at the LHC as a probe of New Physics that couples mainly to third generation of quarks.  We analyze New Physics simplified models with resonances of spin 0, 1 and 2.  The sensitivity of the final states $t\bar t b\bar b$, $t\bar t t \bar t$, $b \bar b b \bar b$ and $t \bar t$ on each of these models is used to identify an important region in parameter space that is still not excluded and where $t\bar t b \bar b$ is the most sensitive final state.  We indicate possible experimental hints and discuss potential issues of observables that rely mainly in Montecarlo predictions.  A new observable is proposed that, at the price of requiring more statistics, reduces the impact of Montecarlo predictions.  We use preliminary 13 TeV results to give a raw estimate on the discovery reach and propose simple improvements on the observables.
\end{abstract}

\vspace*{5cm}
\noindent {\footnotesize E-mail:
{\tt \href{mailto:sequi@df.uba.ar}{sequi@df.uba.ar},
\href{mailto:mestevez@unsam.edu.ar}{mestevez@unsam.edu.ar}}}

\newpage
\section{Introduction}

Since the Standard Model (SM) was coined 50 years ago, searches for New Physics (NP) were influenced by the yearn of a complete theory.  Where a complete theory means that, in addition of solving many observational problems, it also addresses theoretical issues such as ultraviolet (UV) divergences and the hierarchy problem, among others.  Given the negative results in these searches insofar, it could be valuable to begin considering that at the next level of the energy frontier the NP may be hopefully seen only as a partial theory with many patches to be understood and solved in the forthcoming years.  Pushing the possibility of a complete theory, as for instance Supersymmetry \cite{Wess:1974tw,Fayet:1976cr} or Extra Dimensions \cite{Randall:1999vf} among others, for a higher energy level. This new scenario may lead to more phenomenological works.  In this article, motivated by a possible involvement of the third generation of quarks in the Electroweak Symmetry Breaking (EWSB), the large top mass and the small bottom mass, and some experimental hints in final states involving bottom and tops, we propose to further study the final state $t\bar t b \bar b$ as a probe of NP at the LHC.

We pursue the question of what kind of NP would affect mainly this final state and, once this is stated, which region in parameter space is not yet discarded and how this and other observables would affect it.  The answer to the first question requires a new particle that couples mainly --if not exclusively-- to the third generation of quarks, the second part of the question is addressed in the following sections.  The new particle should be singlet in color to avoid interaction with gluons.  If the particle has spin 0, then a Two Higgs Doublet Model (2HDM) \cite{Gunion:1989we} would seem a good theoretical framework; however a 2HDM is a UV-complete framework that pays the price of imposing further conditions for the top and bottom couplings, which is what we want to avoid at this level to keep the analysis as general as possible.  Therefore, we consider scalar or pseudo-scalars 'boson-phobic' \cite{Frederix:2007gi} whose top and bottom couplings have not {\it a priori} restrictions and allows for a more detailed exploration of the parameter space.  A colorless spin 1 particle is what is known as a $Z'$ \cite{Langacker:2008yv} an has less theoretical issues, since it can only couple to right handed fermions.  A spin 2 particle can also couple to only right handed fermions and, as it is shown in this article, it has many other attractive features.

The experimental and theoretical study of final states with heavy flavors has always been an interesting sector where to search for deviations from the Standard Model (SM), and in particular at LHC, Tevatron and LEP this has been specially true for the third generation of quarks. The study of pairs of third generation of quarks has given SM-expected results for $t\bar t$ at the LHC \cite{Carli:2016eiu,Llacer:2015ypa} and Tevatron \cite{Margaroli:2015kxa}, whereas $b\bar b$ is a difficult process to study in these hadronic machines because of the huge background and very few clean results are available at present time \cite{ATLAS:2016gvq,Khachatryan:2015sja}.  However, at LEP the final state $b\bar b$ left one of its larger deviations in what is known as the LEP $b$-anomaly \cite{Olive:2016xmw}.  

The study of final states with three or more third generation quarks at the LHC is a developing field which is generating an important set of new results lately.  The four-bottom final state is usually studied in searches for heavy Higgs which are produced through $b\bar b$ fusion and decays to $b \bar b$ \cite{Mankel:2015ugl,Chatrchyan:2013qga,Khachatryan:2015tra}.  However, this final state has the difficulty of being hidden in QCD background as well as being hard to tag the bottoms.  The four-top final state is many times analyzed as possible signatures of heavy resonances decaying to $t\bar t$ \cite{test0}, and also lately in looking for the SM four-top process \cite{Beck:2016hyi,CMS:2016wig}, which may also hide non-resonant NP \cite{Alvarez:2016nrz}.  Notice, however, that four-top is a very heavy final state and usually its cross-section has important suppressions.  Four-top is also a very populated final state which may yield difficulties in its reconstruction.  The $t\bar t b \bar b$ \cite{CMS:2014yxa,Aad:2015yja} final state is the mid-term between the previous two final states and is very interesting because most of the previously mentioned difficulties are reduced meanwhile still captures the attraction of NP effects. 

Most of the mentioned final states have slight excesses which become larger as the number of $b$-tags increases.  For instance --and not being exhaustive--, in ATLAS four-top search in Ref.~\cite{ATLAS:2016gqb}, the comparison of prediction to data in all control and signal regions yields a systematic increasing deviation as the number of required $b$-tags goes from 2 to 4.  In ATLAS $t\bar t b\bar b$ first study \cite{Aad:2013tua}, the measured fiducial cross-section was 2$\sigma$ beyond the SM prediction.  Later results \cite{Aad:2015yja} including more data and some re-analysis reduced this excess to $1.2\sigma$.  Ref.~\cite{Aad:2015yja} dedicates a full section to discuss possible different tuning of the Montecarlo, mainly on the $g\to b \bar b$ splitting kernels whose tuning has an important effect on the data to prediction agreement.  CMS study on $t\bar t b \bar b$ was from the beginning addressed to reduce the impact of eventual Montecarlo issues due to the $b$-pair in the final state.  In their first work \cite{CMS:2014yxa} they presented a comparison between $\sigma(t\bar t b \bar b)$ and $\sigma(t\bar t jj)$ with a $1\sigma$ excess.  In their second article \cite{Khachatryan:2015mva} they have presented relevant differential measurements in kinematic variables such as $\Delta R_{b\bar b}$ and $m_{b\bar b}$, in which a resonant signal can be distinguished using side-band measurements.  Interestingly, both measurements have $\sim 1.5\sigma$ excesses in the back-to-back bin in $\Delta R_{b\bar b}$ and in the large mass $m_{b\bar b}>170$ GeV bin.  These measurements are also performed in the $t\bar t jj$ final state under similar conditions and no excess is found neither in $\Delta R_{jj}$ nor $m_{jj}$.  It is also interesting to note that they need to apply normalization factors of 2 and 4 for all the predictions on leading and subleading additional $b$ jet quantities to reach agreement with data, reflecting some issue at least in the Montecarlo.

In the last few months preliminary $t\bar t b \bar b$ results from ATLAS \cite{ATLAS:2016btu} and CMS \cite{CMS:2016tlo} appeared on the 13 TeV data.  ATLAS results, before fitting the Montecarlo to data, show again an excess which increases with the number of $b$ tags.  This excess disappears after performing a fitting to data in the background-only hypothesis.  CMS \cite{CMS:2016tlo} total ratio $\sigma(t\bar t b \bar b)/\sigma(t\bar t jj)$ yields a 1.5$\sigma$ excess compared to SM prediction.  However, in this article we will restrict the main analysis to published data on $t\bar t b \bar b$ \cite{CMS:2014yxa,Aad:2015yja}, which corresponds to $7$ and $8$ TeV data.

There are also many theoretical works which have addressed the $t\bar t b \bar b$ final state\cite{Gori:2016zto,Dolan:2016qvg,Jo:2015zxa,Bevilacqua:2014qfa}, most of them as a side analysis.  We focus on this final state in this article and we identify the region in parameter space where its importance is enhanced.

This work is divided as follows. In section \ref{sec:model} we describe the simplified NP models we use along the article.  We also performed a qualitative analysis on the expected behavior of these NP scenarios on the heavy flavour final states.  In section \ref{sec:ttbb} we present our main results, the region in parameter space which has not been probed yet and in which $t\bar t b \bar b$ is expected to be the most sensitive final state.  We discuss some relevant points in section \ref{sec:discussion}, including potential Montecarlo issues present in the $t\bar t b \bar b$ final state.  We propose a new observable which could be useful to reduce the impact of Montecarlo in the distinction of a NP signal, and we perform a raw estimate of the reach of existing observables.  We conclude in section \ref{sec:conclusions}.

\section{Simplified NP models}
\label{sec:model}

Along this section we present four simplified models of NP which couple to the third generation of quarks: a scalar, a pseudo-scalar, a vector, and a graviton.  We restrict our study to electrically neutral NP, although charged NP could also provide interesting NP models.  Depending on the magnitude of the parameters in the model, this kind of NP would be mainly manifest in final states as $t\bar tb \bar b$, $t\bar t t \bar t$, $b\bar b b \bar b$, $t\bar t$ and $b\bar b$.  In particular, the relative sensitivity of these final states to the NP will change drastically depending if the new particle mass is heavier or lighter than $2m_t$.  

To keep the discussion as general as possible, we make no assumptions in the underlying model, we focus on the phenomenology and present possible simplified NP Lagrangians which couple to top and bottom quarks.

\subsection{Scalar $\phi$}

The case of a massive scalar field that couples exclusively to top and bottom is one of the simplest NP models.  However, a scalar field has potential issues, since involves interactions with Left and Right quarks, which yields a vertex with a non-trivial SM EW gauge structure.  Nevertheless, this kind of interacting terms can be achieved in a variety of models without stronger complications, since the effective vertex can be understood, for instance, as coming from higher dimensional operators or from a spontaneously breaking of the symmetry.   The parameters of such a model can be in general matched to UV complete models such as some of the many available Two Higgs Doublet Model (2HDM) (See for instance, and not being exhaustive, Refs.~\cite{Craig:2013hca,Branco:2011iw,Gunion:1989we}). The main issue here may be how independent can be the top and bottom couplings.  In a 2HDM this is addressed with the free parameter $\tan \beta$, although it could be the case that for given arbitrary values of the top and bottom couplings this particular UV completion scenario may be constrained in the parameter space from other experimental limits placed within the specific 2HDM case.  In any case, we stress that the analysis in this work is focused in the phenomenological NP Lagrangians, rather than in UV complete theories.  Another problem with this model is that for large couplings to bottom quarks and small masses, it could enter into conflict with LEP precision observables.

For the scalar case we study the following simplified Lagrangian
\begin{equation}
{\cal L}_{\phi}^{tree} = - c_{\phi t} \, \bar t_L \phi t_R -  c_{\phi b} \, \bar b_L \phi b_R + h.c.
\end{equation}

A one-loop effective coupling to gluons $gg\phi$ can be added to the Lagrangian due to a top and bottom loop. The extra Lagrangian in this case reads
\begin{equation}
{\cal L}_{\phi}^{loop} = 
\frac{\alpha_s}{12\sqrt{2}\pi} \left[ \frac{c_{\phi t}}{m_t} F(z_t) + \frac{c_{\phi b}}{m_b} F(z_b)\right] \phi \, G^a_{\mu\nu} G_a^{\mu\nu} 
\end{equation}
where $$z_{t,b} = (2 m_{t,b}/M)^2$$ with $M$ the resonance mass (in this case $M=M_\phi$), and $F(z)$ is the loop function that can be found in the Appendix.

The NP interacting effective Lagrangian is therefore
\begin{equation}
{\cal L}_{\phi} = {\cal L}_{\phi}^{tree} + {\cal L}_{\phi}^{loop}.
\end{equation}

\subsection{Pseudoscalar $A$}

The pseudoscalar case, which could also be understood as a member of a 2HDM, has the following Lagrangian
\begin{equation}
{\cal L}_{A}^{tree} = - c_{A t} \, \bar t_L A \gamma^5 t_R -  c_{A b} \, \bar b_L A \gamma^5 b_R + h.c.
\end{equation}

The one-loop effective Lagrangian that couples $ggA$ reads
\begin{equation}
{\cal L}_{A}^{loop} = 
\frac{\alpha_s}{8 \sqrt{2} \pi} \left[ \frac{c_{A t}}{m_t} H(z_t) + \frac{c_{A b}}{m_b} H(z_b)\right] A \, G^a_{\mu\nu} \tilde G_a^{\mu\nu} .
\end{equation}
Where the loop function $H(z)$ can be found in the Appendix.  The full pseudoscalar Lagrangian reads ${\cal L}_{A} = {\cal L}_{A}^{tree} + {\cal L}_{A}^{loop}$.

\subsection{Vector Z'}
A spin one, colorless and neutral $Z'$ with mass $M_{Z'}$ which couples exclusively to $t$ and $b$ ends up being the model with less potential issues.  To keep the top and bottom couplings as free parameters independently of the SM EW gauge group, we keep only couplings to right handed quarks. This also helps to avoid conflicts with LEP precision measurements. The interaction of such a $Z'$ with the SM particles reads
\begin{equation}
{\cal L}_{Z'} = - c_{Z't} \, \bar Z'_\mu t_R \gamma^\mu t_R - c_{Z'b} \, Z'_\mu \bar b_R \gamma^\mu b_R.
\end{equation}
Since $Z'$ has spin one, it cannot couple at any order to a $gg$ initial state.
	
\subsection{Graviton G} 

We consider an effective Lagrangian for a spin-2 graviton with field $\hat G_{\mu\nu}$.  The tree-level interaction Lagrangian reads
\begin{equation}
{\cal L}_{G}^{tree} = - \frac{i}{2 \Lambda} \hat G ^{\mu\nu} \left[ c_{Gt} \left( \bar t_R \gamma_\mu \overset{\text{\tiny $\leftrightarrow$}}{D}_\nu t_R - \eta_{\mu\nu} \bar t_R \gamma^\rho \overset{\text{\tiny $\leftrightarrow$}}{D}_\rho t_R \right) + c_{Gb} \left( \bar b_R \gamma_\mu \overset{\text{\tiny $\leftrightarrow$}}{D}_\nu b_R - \eta_{\mu\nu} \bar b_R \gamma^\rho \overset{\text{\tiny $\leftrightarrow$}}{D}_\rho b_R \right) \right] .
\end{equation}
Where $\bar f \gamma_\mu \overset{\text{\tiny $\leftrightarrow$}}{D}_\nu f = \bar f \gamma_\mu D_\nu f - D_\nu \bar f \gamma_\mu f$. For the purposes of this work, we only use the partial derivative term in the covariant derivative, since we neglect the 4-particle interaction contained in the other terms.

Contrary to the previous NP models, the spin-2 Lagrangian needs dimensional couplings, and this is why the dimensional constant $\Lambda$ in the denominator.  Along the remaining of this article we set $$\Lambda=1\ \mbox{TeV}.$$  This arbitrary choice matches the cross-sections $\sigma(b\bar b \to G)=\sigma(b \bar b \to \phi)$  for the same numerical values of the couplings $c_{\phi b} = c_{X b}=1$ and $M_G=M_\phi= 300$ GeV.  Of course that this agreement is lost as the numerical values of the couplings or the mass of the resonance are modified, since both models have a different dynamic.  This choice is just to have a contact point for the spin-2 NP model whose couplings are not dimensionless.

The one-loop effective Lagrangian that couples $ggG$ reads
\begin{equation}
{\cal L}_{G}^{loop} = 
-\frac{\alpha_s}{12\pi \Lambda} \left[ c_{G t} A_G(z_t,\mu_0) + c_{G b} A_G(z_b,\mu_0) \right] \hat G_{\mu\nu} \left( \frac{\eta^{\mu\nu}}{4} \, G^a_{\rho\sigma} G_a^{\rho\sigma} - G^{a\mu}_\rho G_a^{\nu\rho} \right) .
\end{equation}
Here $\mu_0$ is the renormalization scale and the loop function $A_G (z,\mu_0)$ can be found in the Appendix.  The full spin-2 Lagrangian reads ${\cal L}_G = {\cal L}_{G}^{tree} + {\cal L}_{G}^{loop}$.

\subsection{General features}

All four simplified models presented in the previous paragraphs have essentially the same Feynman diagrams.  Depending on the point in parameter space under study, each diagram will have different importance as a function of the available energy in the proton PDF's and the phase space available for the possible decays. 

Notice that for simplicity we have included in the loop only top and bottom quarks, and we will restrict to real and same-sign couplings.  In principle, complex combination of these couplings and/or new heavy NP resonances in the loop function, can modify the effective vertex of the NP to two gluons.

\begin{figure}[!htb]
\begin{minipage}[b]{0.49\textwidth}
\begin{center}
\includegraphics[width=0.9\textwidth]{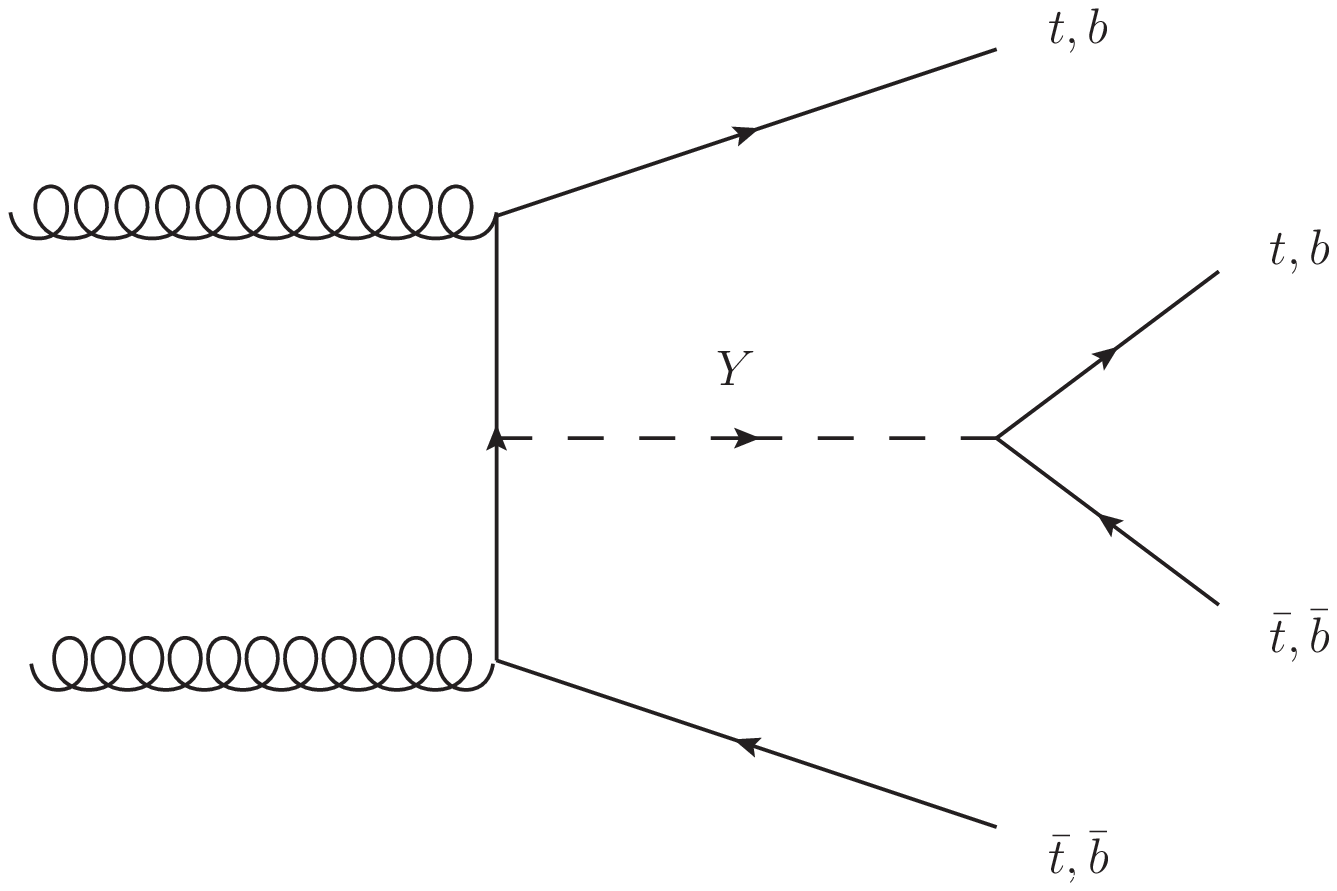}
\newline
\small{D1}
\end{center}
\end{minipage}
\begin{minipage}[b]{0.49\textwidth}
\begin{center}
\includegraphics[width=0.9\textwidth]{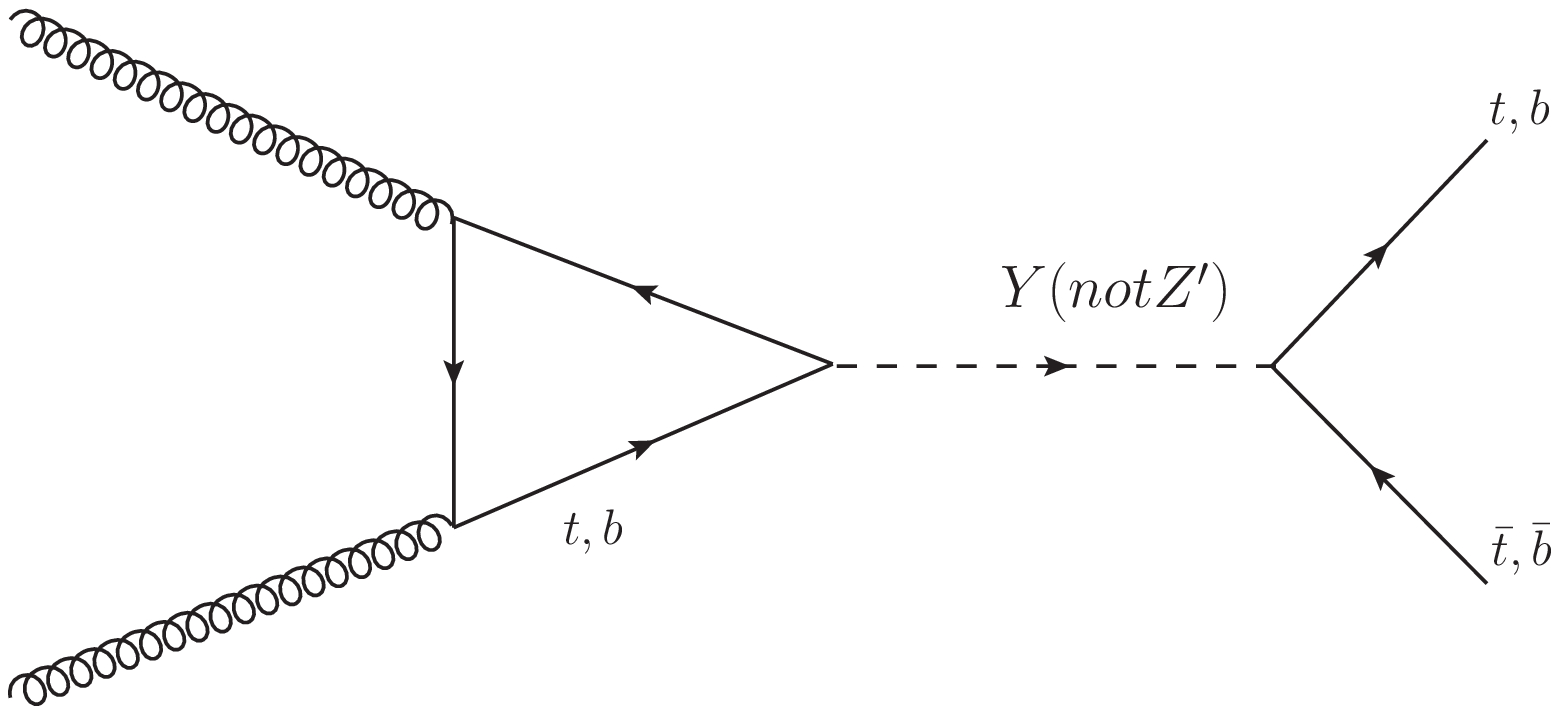}
\newline
\vskip .3cm
\small{D2}
\end{center}
\end{minipage}
\newline
~\vskip .5cm
\begin{minipage}[b]{0.49\textwidth}
\begin{center}
\includegraphics[width=0.9\textwidth]{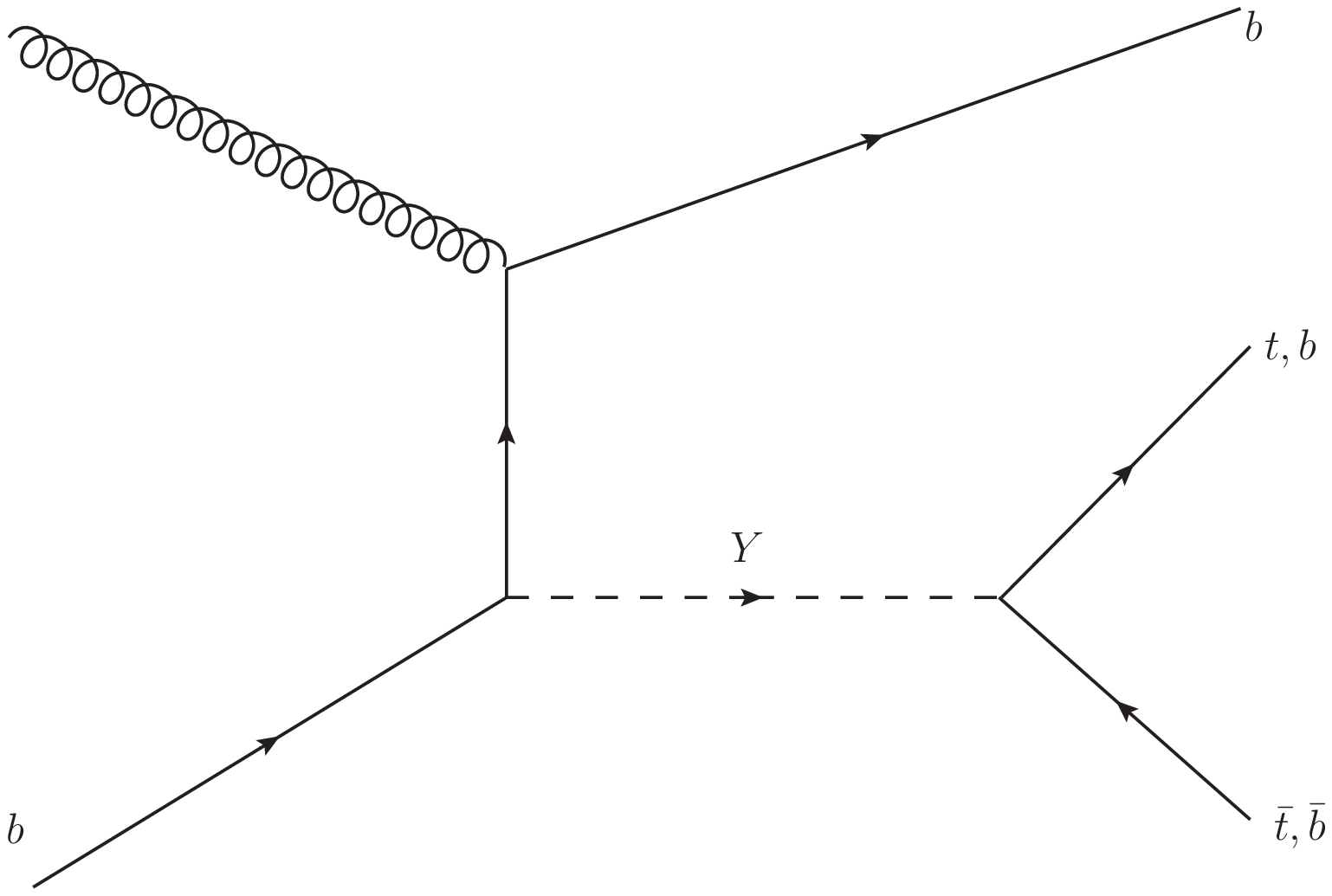}
\newline
\small{D3}
\end{center}
\end{minipage}
\begin{minipage}[b]{0.49\textwidth}
\begin{center}
\includegraphics[width=0.9\textwidth]{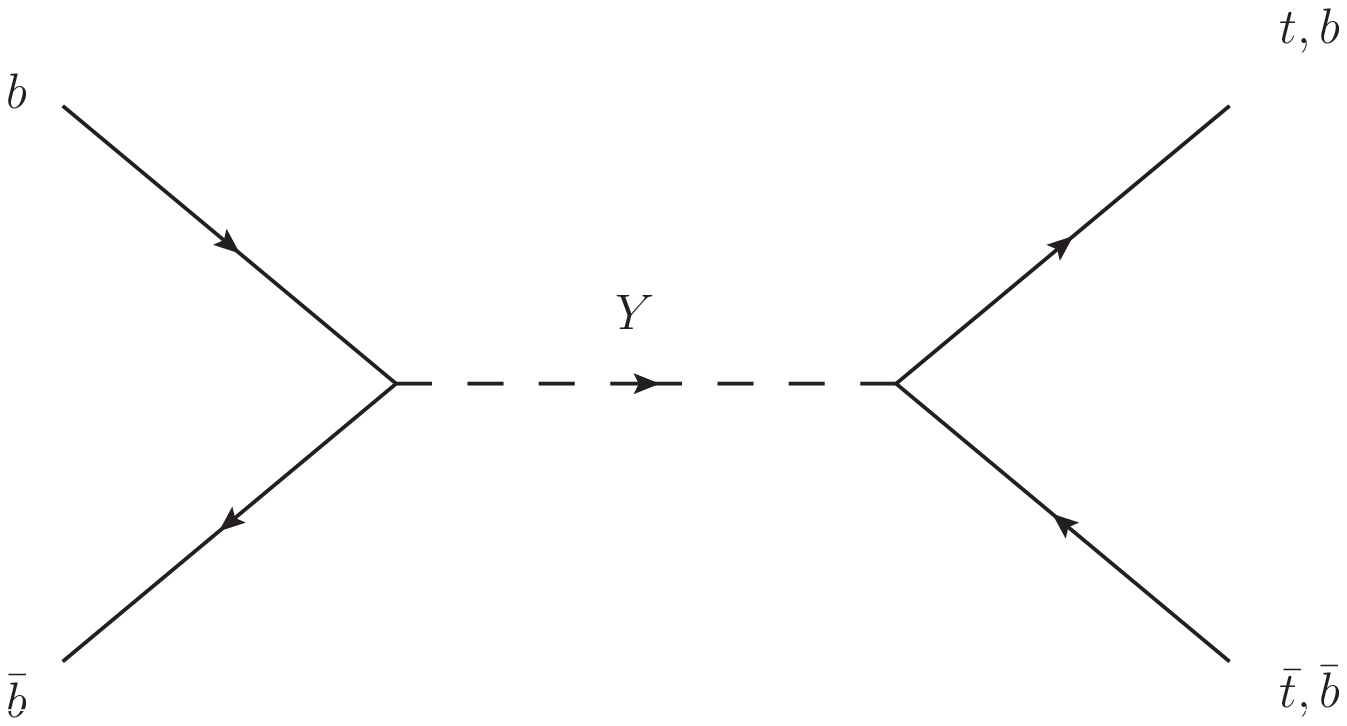}
\newline
\vskip .2cm
\small{D4}
\end{center}
\end{minipage}
\vspace{.5cm}
\newline
\caption{Relevant Feynman diagrams for processes in which NP that couples to the third generation of quarks could have an important effect in final states with 2 or more heavy quarks.  $Y$ refers schematically to any of the possible new particles described in text: $\phi$, $A$, $Z'$  or $G$; except in D2 where $Z'$ is forbidden because of Yang's theorem.}
\label{diagrams}
\end{figure} 

We show in Fig.~\ref{diagrams} the NP relevant diagrams.  We call $Y$ to any of the possible new particles in the previous sections ($\phi$, $A$, $Z'$ or $G$) and $M$ to any of their masses; except where explicitly stated different. 

Diagrams D1, D3 and D4 --which are in general the most important production mechanism for $Y$--, are essentially the same diagram when the internal line in D1 is a $b$, with the only difference of whether the initial $b$ quark is assumed in the proton, or not.  As a matter of fact, the real difference in assuming the $b$ in the PDF or not in each one of these processes depends on the factorization scale and the momentum of the $b$ quark.  For the phenomenological analysis that follows, we refer to these diagrams as whether the other $b$ that comes from the corresponding gluon splitting is detected or not.  In this sense, in addition to the $Y$ decay products, we consider diagrams D1, D3 and D4 to have 2, 1 and none extra $b$'s, respectively.  This would make an important effect, since integrating out a $b$ yields an inclusive cross-section larger than the exclusive one; specially because of the forward ($\eta>2.5$) integration.  Although in principle D4 would be the process with the more inclusive cross-section, its background $pp\to t \bar t$ is the largest of all and therefore its relevance is diminished.  However, when comparing D1 and D3, they share the same main background, but D3 is more inclusive.  This yields the important conclusion that, if the full final state is not required to be reconstructed, is in general better to look for the final states\footnote{Along this work we refer indistinctly to $t\bar tbX$ or $t\bar t \bar bX$.  Analogously for $b\bar b bX$ and $b\bar b \bar bX$.} $t\bar tbX$ and $b\bar b bX$ rather than $t\bar tb\bar b$ and $b\bar b b \bar b$, respectively; where $X$ refers to the inclusiveness of the process.   This last statement is however not totally true for $M < 2m_t$, since in that case we do not expect an improvement in $t\bar t bX$ with respect to $t\bar t b \bar b$, although we do in $b\bar b bX$ with respect to $b\bar b b\bar b$.  This is because $Y$ cannot decay to $t\bar t$ ($Y \not \to t \bar t$) and therefore the $b\bar b$-pair in $t\bar t b \bar b$ must come from $Y$, thus integrating out one $b$ which is not expected forward should not make an important difference.  This observation has important consequences in the phenomenology studied in the following sections: for $M<2m_t$ the final state $b\bar b b X$ will be the most sensitive channel in a larger region in parameter space.

An important observation regarding diagram D1 and conservation of angular momentum is to be noticed depending whether $Y$ has spin 0, 1 or 2.  Assuming null orbital angular momentum in final state, the case $Y=\phi$ or $A$ yields a final state whose total angular momentum can only be 0 (since the initial state cannot have total angular momentum 1).  Whether the case $Y=Z'$ yields a final state whose total angular momentum can be either 0 or 2.  Since the initial state can couple in total angular momentum 0 or 2, one concludes that the case $Y=\phi$ or $A$ is suppressed because of conservation of angular momentum in comparison to the $Y=Z'$ case\footnote{A similar and interesting situation holds in $pp\to t\bar t h$, which is so angular-momentum-suppressed that adding a final gluon, $pp\to t \bar t h g$ ($p_T(g)>20$ GeV and $|\eta(g)|<5$), yields a similar cross-section because of the extra angular momentum channel.  As a matter of fact, $pp\to t\bar t h$ ends up having a considerable fraction of $q\bar q$ initial state collision to relieve this angular momentum tension.}. In a similar way, for $Y=G$ the final state can only have total angular momentum $2$, and therefore cannot come from an initial state with total angular momentum $0$.  We conclude from these observations that $Y=Z'$ is relatively much easier to discard, whereas $Y=\phi,\,A$ and $G$ are harder to explore.  This is explicitly seen in next section.

Diagrams D2 and D4 (with no extra $b$) in Fig.~\ref{diagrams} yield a resonant final state in $t\bar t$ or $b\bar b$.  Diagram D2 interferes destructively with the SM process $gg \to t\bar t/b \bar b$ \cite{Hespel:2016qaf} and therefore its usefulness needs to be explored carefully.  The $t\bar t$ final state is the only one studied with relevant accuracy, but only for resonances up to $\Gamma_Y/M$ larger than a few percents (and not much more, since wide $t\bar t$ resonance searches have not been explored in depth \cite{Hespel:2016qaf}).   Given the precision in the results in $t\bar t$ invariant mass spectrum bump searches \cite{Aad:2015fna}, this diagram could in principle be relevant for our purposes.  On the other hand, diagram D4 is in general explored in extra scalar SUSY searches \cite{Chatrchyan:2013qga,Mankel:2015ugl}, however always an extra $b$ is required and one finishes in diagrams D1 and D3.   It is interesting to notice that a $b\bar b$ resonance search with no extra $b$  --which would correspond to D2-- is a difficult task due to large background for small $M$ and to the difficulty in tagging $b$-jets for large $M$ \cite{Khachatryan:2015sja,ATLAS:2016gvq}.  It is worth observing that a new resonance with $M<2m_t$ and large couplings to tops would be detectable through processes as D1 and D2, in both cases with $Y\to b\bar b$.  D1 would be suppressed due to phase space and D2 due to the loop.  

The limit of a heavy $Y$ ($M \gtrsim 1$ TeV) it is interesting for the $t\bar t bX$ final state.  In fact, since very hard $b$'s have a small tagging efficiency \cite{Khachatryan:2015sja,ATLAS:2016gvq}, in this limit it is convenient to have the $b$ quarks coming from the gluon-splitting in D1 and the new particle decaying to top quarks, $Y\to t\bar t$.  Moreover, very hard tops are easy to tag using boosted top tagging techniques.  Notice also that in this limit the $b\bar b b X$ final state will not be as useful because of the same issue in tagging hard $b$'s.

In addition to the above discussion, there are other diagrams which could be important in special cases.  For instance, a light ($M<2m_t$) NP that couples exclusively to top quark would be visible in a $t\bar t t \bar t$ final state and $t$-channel diagrams would affect the total cross-section with no other special features, as studied in Ref.~\cite{Alvarez:2016nrz}.  Also, for large coupling to $b$-quarks and $M \gtrsim 2m_t$, the $t$-channel $b\bar b \to YY$ diagram is important, and some times more important than $gg$-fusion.  At larger $M$ this diagram becomes suppressed because of the PDF's.

\section{ttbb sensitivity to NP}
\label{sec:ttbb}

In this Section we perform a quantitative comparison of the sensitivity of different channels to the NP presented in Sect.~\ref{sec:model}.  We compare the process $pp \to t\bar t b \bar b$ to the other processes where this NP would manifest itself: $pp \to t\bar t t \bar t,\ b \bar b b \bar b$ and $ t \bar t$. We do not consider $pp\to gg/\gamma\gamma$ since they are doubly loop-suppressed in production and decay, however other final states such as $\gamma\gamma t\bar t/b \bar b$ could be interesting to study in the future even though they are loop-suppressed in the decay.

A study of the effects of the NP in the above final states could be performed in at least two different ways.

One possibility is to design search strategies for each one of the final states, and compute the exclusion/discovery reach for each final state, to conclude which channel is the most sensitive and which luminosity is required to exclude/discover each point in parameter space.  

A second and more practical path is to use the already available search strategies used by the experimentalist whose sensitivity comes summarized in the 95\% C.L.~limits on the cross-section ($\sigma$) times acceptance ($A$) times branching ratios ($BR$) of an hypothetical new particle.    Using these results, one can compute in a given channel the ratio of the expected NP $\sigma\times A\times BR$ to the corresponding expected limit in those search strategies.  This ratio is the strength of the given channel expressed in such a way that it can be directly compared to the same ratio in other channels.  That is, they are in the same units of corresponding sensitivity. More precisely, we define the strength of a given final state as  
\begin{equation}
{\cal S} = \frac{\sigma^{NP} \times BR \times A}{\sigma^{lim} \times BR \times A}.
\label{strength}
\end{equation}
Where $\sigma^{NP}$ is the NP cross-section, and $\sigma^{lim}\times BR \times A$ is the 95\%C.L. experimental limit expected in the given channel.  When the experimental limit is differential in some variables, $\sigma^{lim}$ results dependent on the given point in parameter space.  To compare the strength ${\cal S}$ of the different channels, one should take experimental limits with the same center of mass energy and luminosity.  In this case, we have unified the limits to a center of mass energy of $\sqrt s = 8$ TeV and a luminosity of 20 fb$^{-1}$.  This procedure has been used in Refs.\cite{HP2,Alvarez:2016ljl}, and is explained in full detail in Ref.~\cite{Alvarez:2016ljl}.  In this last work it is discussed and shown that the analysis of the most sensitive final state in a given point in parameter space is approximately equally valid for different center of mass energies as long as the search strategy is not drastically changed and the production processes are not significantly modified due to changes in the PDF's.  In this sense, we adopt in this article this second option: we study the most sensitive final state in available search strategies and propose that this relationship is approximately valid for LHC Run II.  If one assumes that experimental limits in the studied final states have a similar scaling with luminosity, then the channel with larger ${\cal S}$ would be the first one to observe or exclude the postulated NP. 

We have selected the most sensitive available published searches for $8$ TeV and $20$ fb$^{-1}$ in the final states $t\bar tbX$, $b \bar b b X$, $t \bar t t \bar t$ and  $t\bar t$, and compare the sensitivity in these searches to the proposed NP as a function of the parameter space.  In all cases we have predicted the excess due to NP using MadGraph 5 \cite{Alwall:2014hca,Alwall:2011uj}.
\begin{figure}[ht!]
\centering
\includegraphics[width=.8\textwidth]{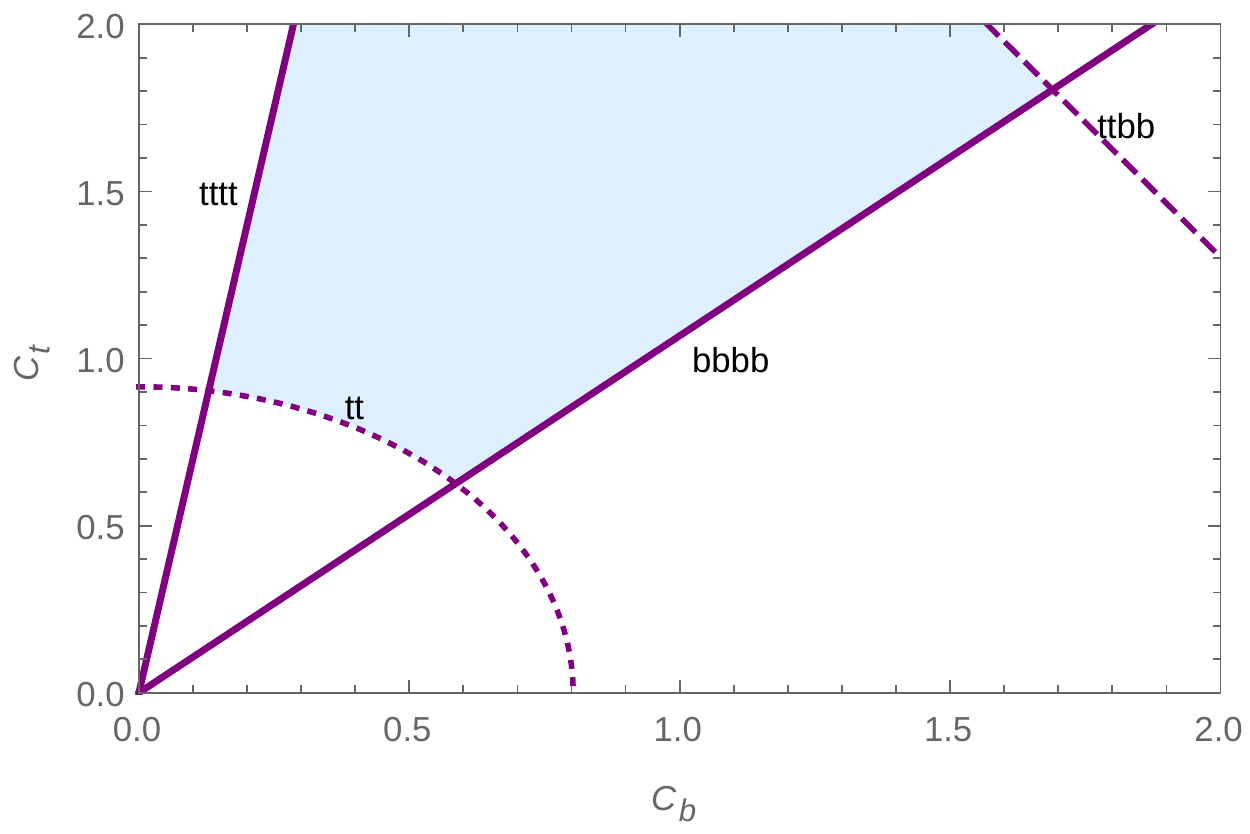}
\caption{
\small Reference plot for region in parameter space most sensitive to $t\bar t b X$.  Solid lines represent borders in which another observable is more sensitive, although those points are not discarded.  Long-dashed represents regions where ${\cal S}(t\bar t b X)=1$ and therefore points beyond that limit are discarded.  Short-dashed represents the $t\bar t$ limit in which the resonance width goes above $8\%$ of its mass and the $t\bar t$ resonance search looses validity \cite{Hespel:2016qaf}.
}
\label{main}
\end{figure}

\begin{figure}[ht!]
\begin{minipage}[c]{0.50\textwidth}
\includegraphics[width=\textwidth]{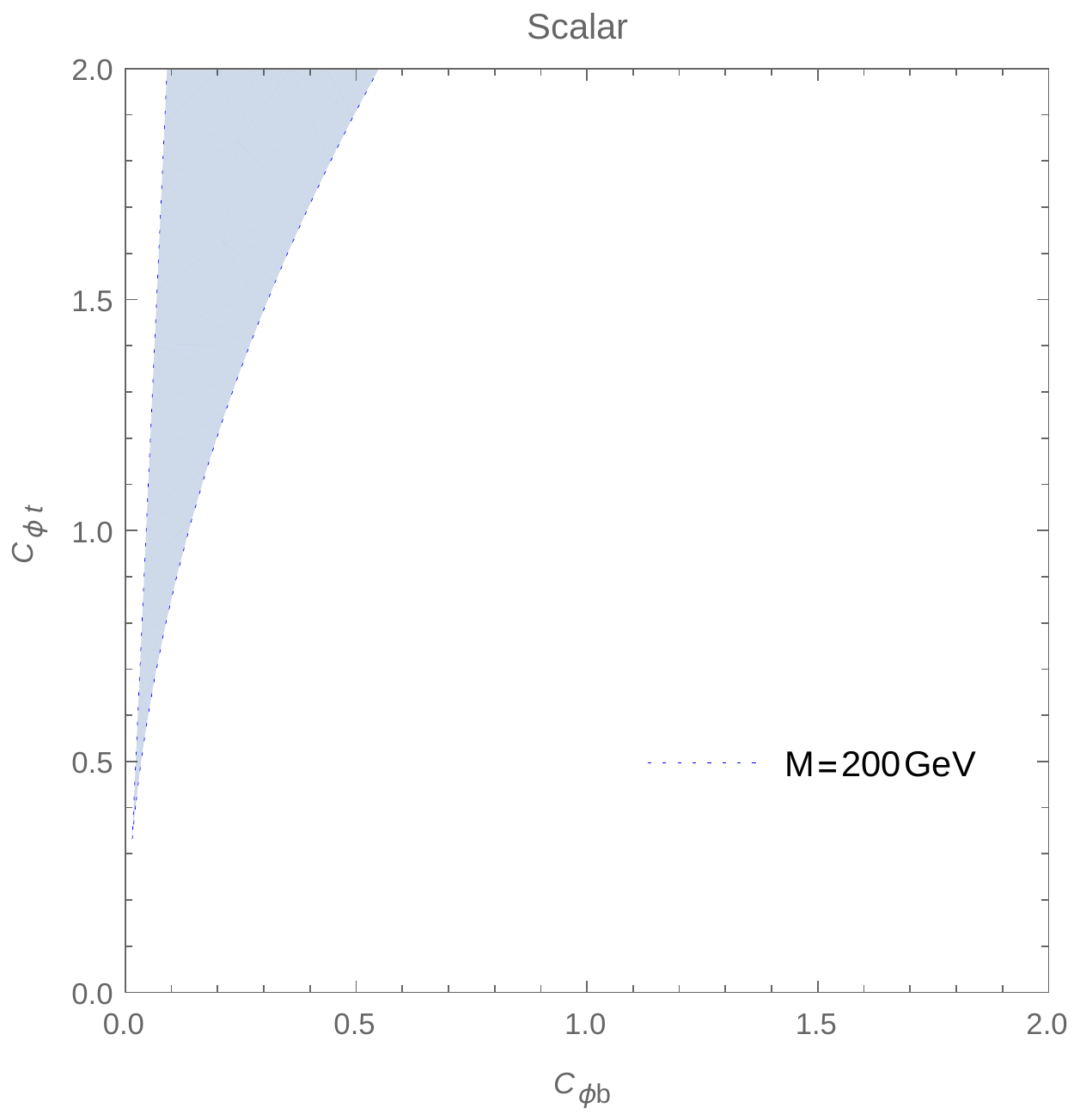}
\end{minipage}
\begin{minipage}[c]{0.50\textwidth}
\includegraphics[width=\textwidth]{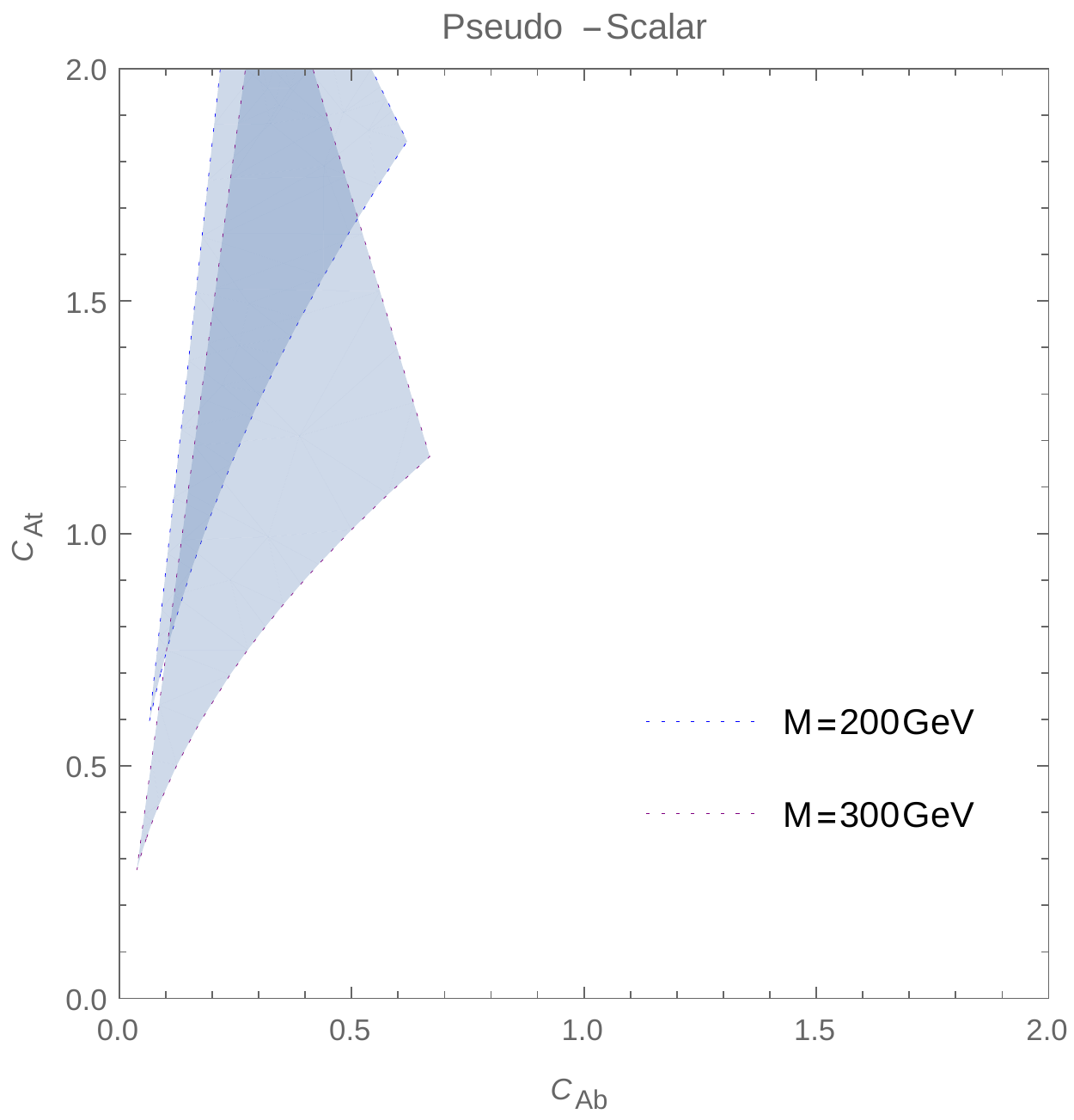}
\end{minipage}
\\
\begin{minipage}[c]{0.50\textwidth}
\includegraphics[width=\textwidth]{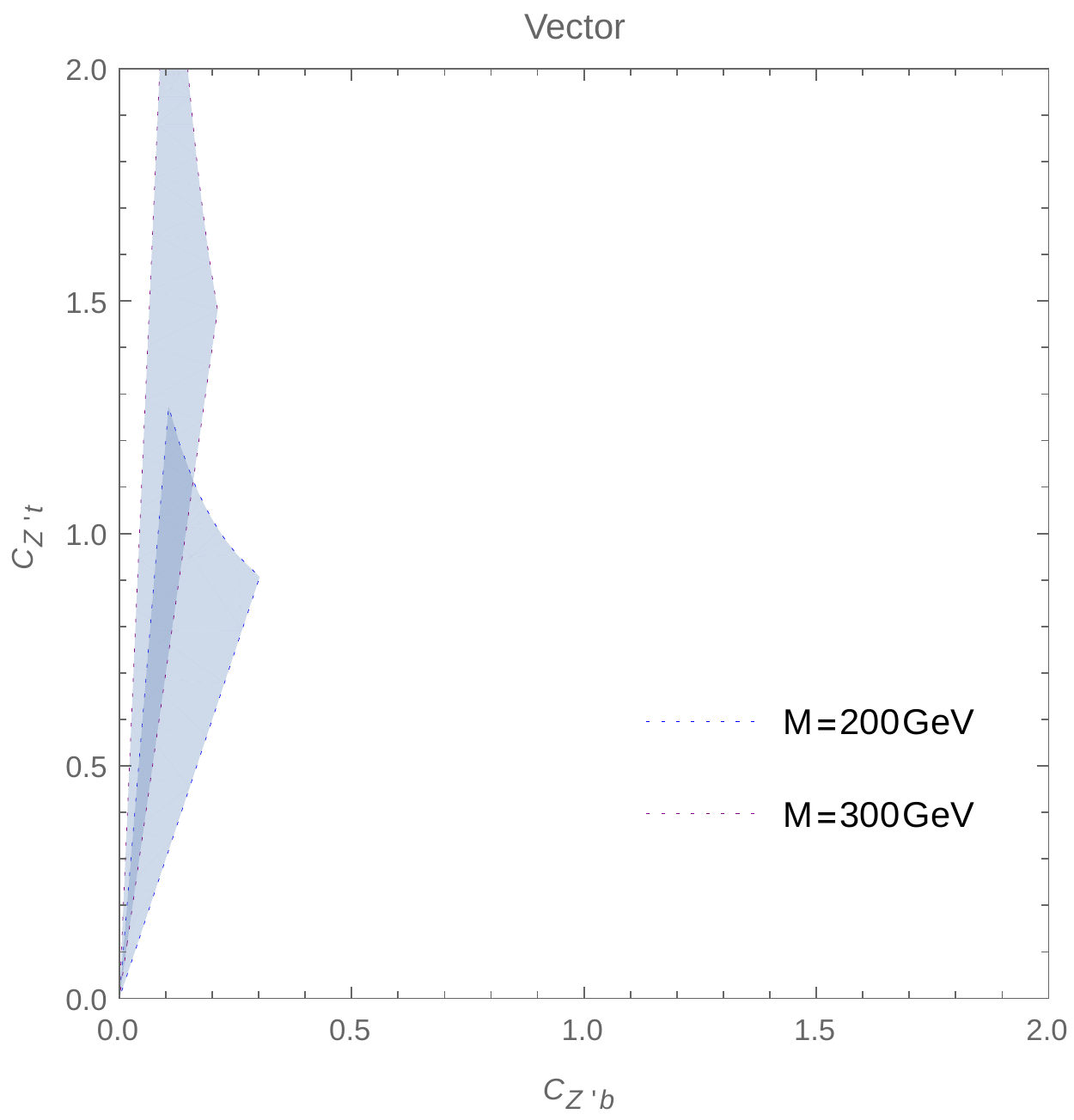}
\end{minipage}
\begin{minipage}[c]{0.50\textwidth}
\includegraphics[width=\textwidth]{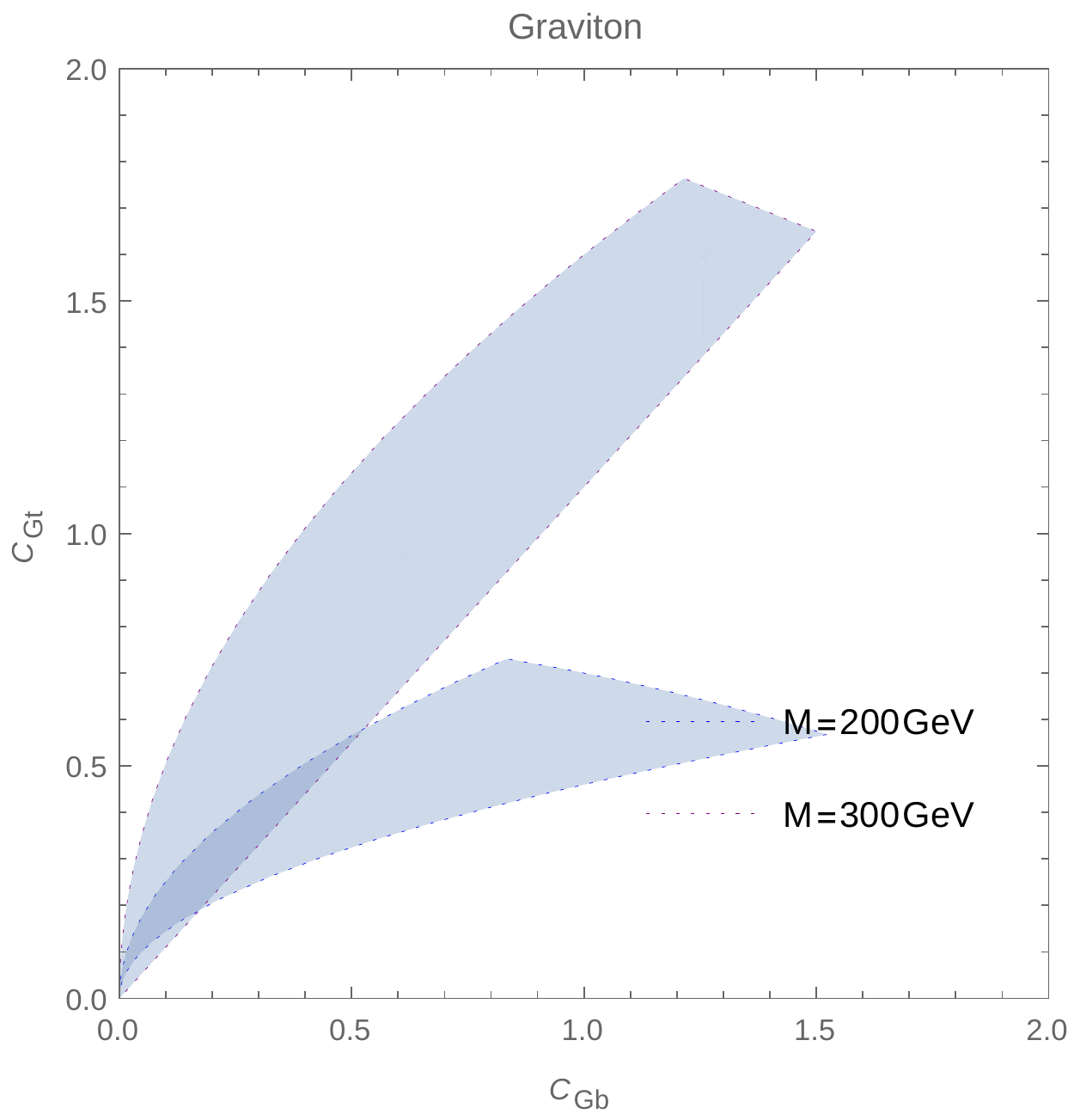}
\end{minipage}
\caption{
\small 
$t\bar t b X$ most sensitive region for small masses  $M < 2 m_t$.  Since the $gg\to b\bar b Y,\ y\to t\bar t$ process is not allowed in this region, the $b\bar b b X$ final state has a large domination in sensitivity. See text for details.
}
\label{Msmall}
\end{figure}

For the final state $t\bar t b X$ we have worked with the search reported by ATLAS in Ref.~\cite{Aad:2015yja}.  To enhance the sensitivity to our NP we have chosen the reported measurement in $t \bar t$ with at least one additional $b$-jet, since this allows the other $b$-jet to be forward and not necessarily detected.  This measurement, as all $t\bar t b X$ results, contain large systematic uncertainties due to the difficulties in modeling QCD effects and in tagging the bottom quarks.  Following Ref.~\cite{Aad:2015yja} and assuming no NP contributions in their measurement, we get that a 95\% C.L. sensitivity corresponds to a cross-section 53\% of the SM expected in the defined fiducial cross-section.  An improvement in the sensitivity of this final state is expected to be achieved when differential measurements use data driven methods.  CMS results in Ref.~\cite{Khachatryan:2015mva} work in that direction, however their sensitivity does not improve ATLAS yet because of the still little number of reconstructed events.

One of the most sensitive final states for the proposed NP is $b \bar b b X$.  We extract the 95\% C.L. sensitivity limits in this final state from Ref.~\cite{Khachatryan:2015tra}.  This work looks for an excess in the invariant mass of the leading $b$'s.  Since the binning in the invariant mass ranges from $\sim 15\%-50\%$ of the sought mass, and we have verified numerically that NP-SM interference effects are not important, we take the limits from this work regardless of the resulting width of the resonance.

The sensitivity in the final state $t \bar t t \bar t$ should be extracted from a search that does not assume resonance production other than through interaction with top quark.  According to \cite{Alvarez:2016nrz}, the closer to this is the SM $t \bar t t \bar t$ production.  We therefore use the 95\% C.L.~limits found in Ref.~\cite{Khachatryan:2014sca} of $32$ fb as the limit cross-section for this final state.

The sensitivity in the channel $t\bar t$ has been extracted from Ref.~\cite{Aad:2015fna}, however we have followed Ref.~\cite{Hespel:2016qaf} in what respects to signal-background interference in  $t\bar t$ resonance searches.  In particular, following their procedure, we have discarded this limit for points in parameter space where $\Gamma/M>8\%$.  We have simulated SM+NP to take into account the interference effects and, where comparable, we reproduce the results in Ref.~\cite{Hespel:2016qaf}.

Using the $\sigma^{lim}$ extracted from the searches described above, unified to $\sqrt s = 8$ TeV and $L=20$ fb$^{-1}$ (see \cite{Alvarez:2016ljl}), and using the predicted signal cross-section times branching ratios times acceptance for each one of these searches, we have computed the strength ${\cal S}$ for all the cases and compare in which points in parameter space is ${\cal S}(t\bar t b X)$ greater than all others.  In these points we expect the  $t\bar t b X$ final state to be the most sensitive channel to the NP described in Sect.~\ref{sec:model}.  Also, in the points in parameter space in which any ${\cal S} > 1$, we consider that point to be discarded.

We have scanned the NP parameter space in different masses going from $M=200$ GeV to $M=3$ TeV, and for the numerical couplings $c_{ij} \in [0,2]$ ($i=\phi,A,Z',G$ and $j=t,b$).  We present the results of the regions in parameter space in which $t\bar t bX$ is the most sensitive final state, and understand which is the most sensitive final state in each border.

\begin{figure}[ht!]
\begin{minipage}[c]{0.50\textwidth}
\includegraphics[width=\textwidth]{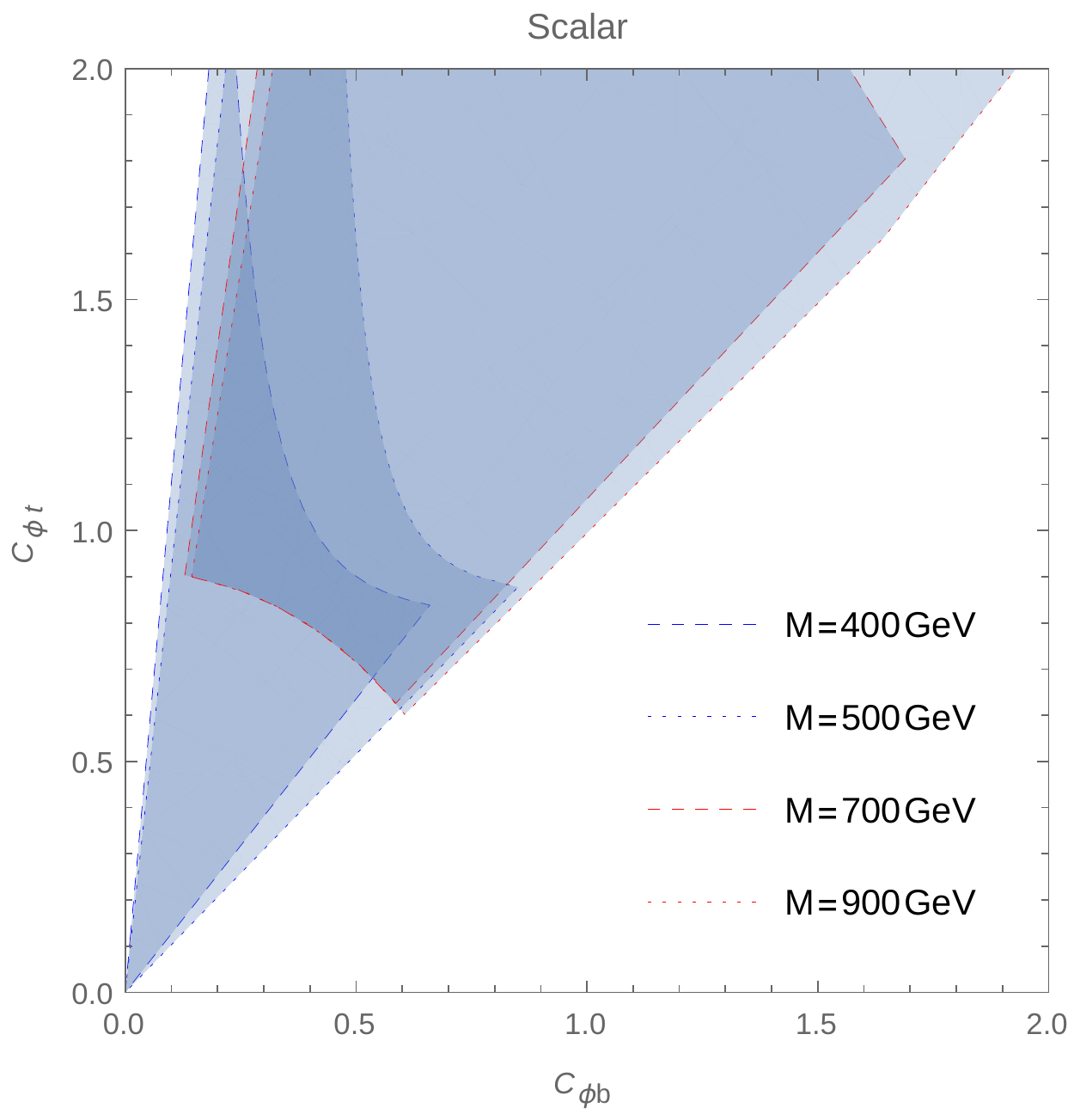}
\end{minipage}
\begin{minipage}[c]{0.50\textwidth}
\includegraphics[width=\textwidth]{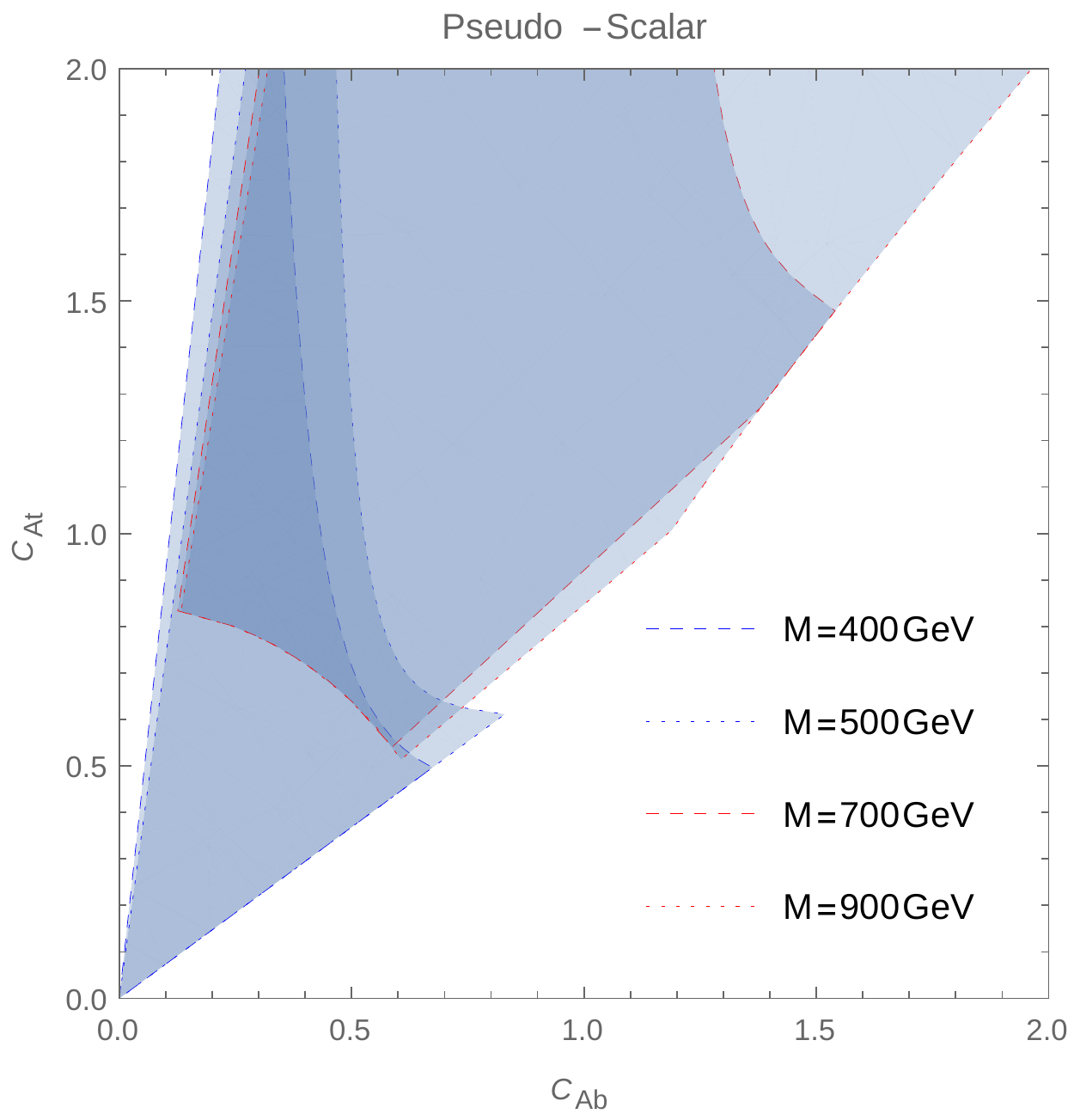}
\end{minipage}
\\
\begin{minipage}[c]{0.50\textwidth}
\includegraphics[width=\textwidth]{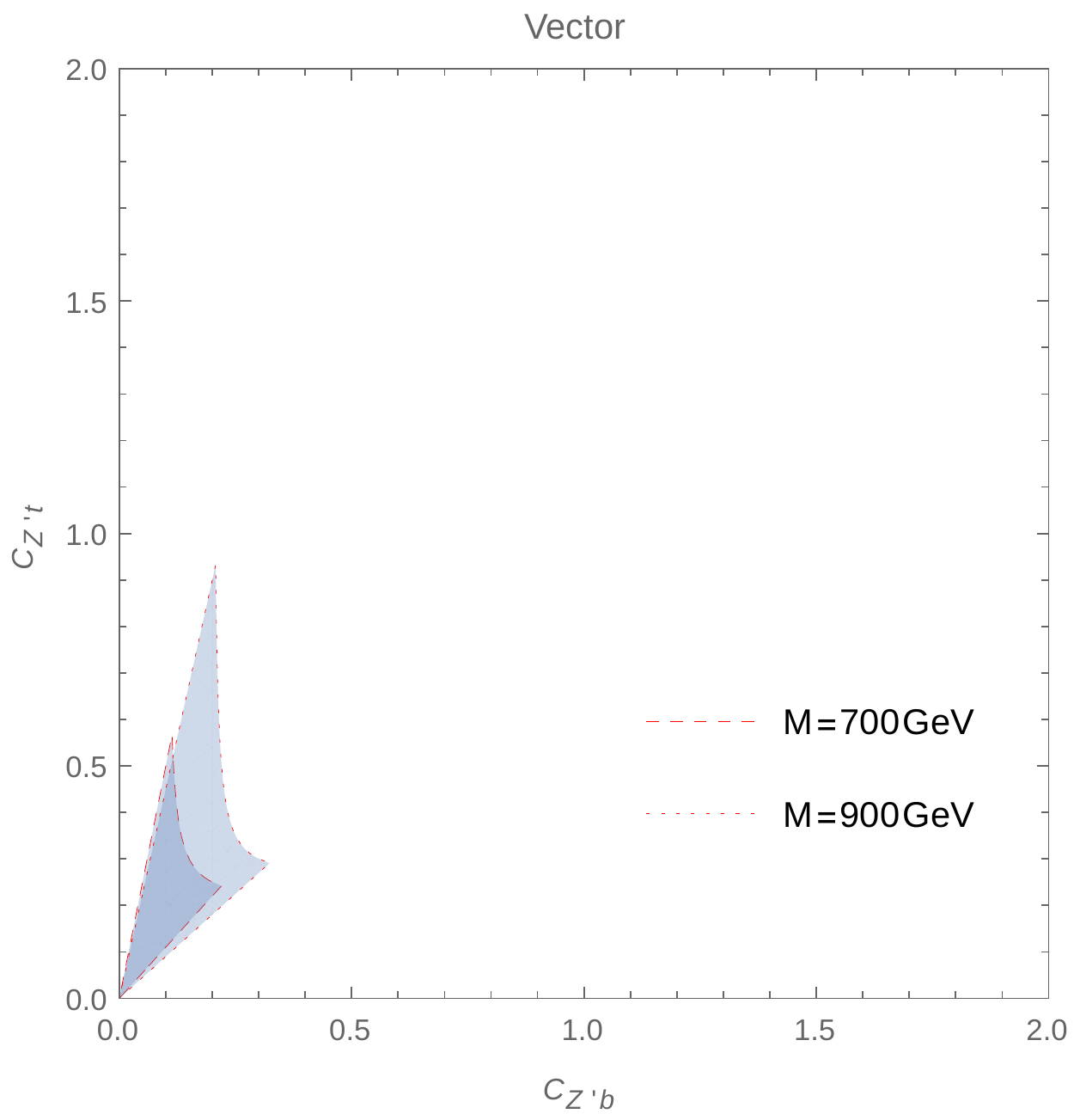}
\end{minipage}
\begin{minipage}[c]{0.50\textwidth}
\includegraphics[width=\textwidth]{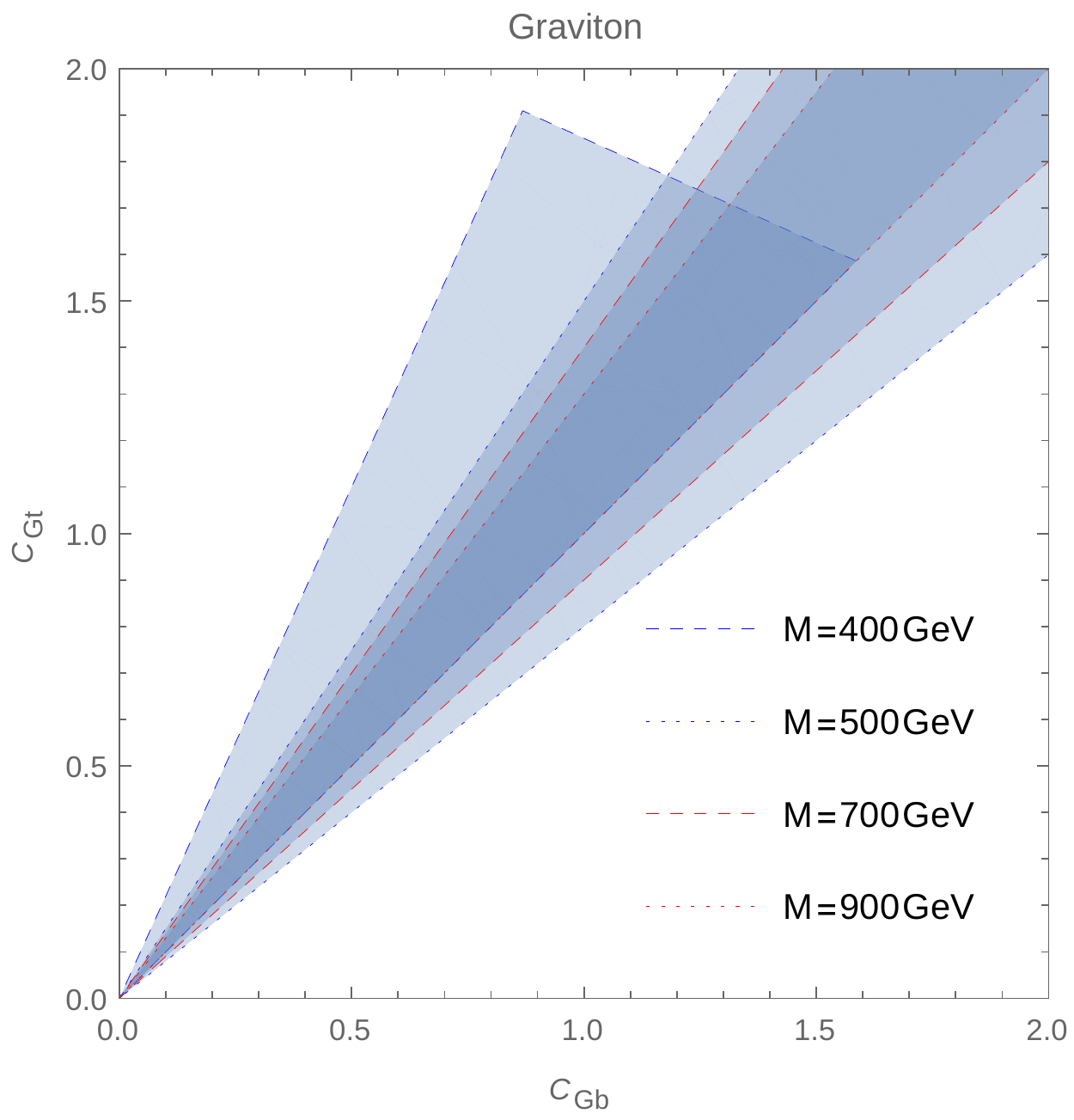}
\end{minipage}
\caption{
\small
$t\bar t b X$ most sensitive region for intermediate masses $2m_t < M < 1$ TeV.  In this case the $b\bar b b X$ domination gets balanced by the $t\bar t b X$ final state.   Also a large region is discarded for the spin-1 $Z'$ resonance due to angular momentum considerations.  See text for details.
}
\label{Mmedium}
\end{figure}

In Fig.~\ref{main} we plot in detail for an arbitrary NP case how the $t\bar t bX$ most sensitive region looks, and which are the observable that constrain this region.  As expected, the right side of the most sensitive region is usually constrained by the $b\bar b b \bar b$ observable, which becomes sensitive as the bottom coupling increases; here ${\cal S}(b\bar b b X)> {\cal S}(t\bar t b X)$.  Analogously, the upper left side is constrained by the $t\bar t t \bar t$ final state: ${\cal S}(t\bar t t \bar t)> {\cal S}(t\bar t b X)$.  On the up right side, there is a constraint that corresponds to direct exclusion by the $t\bar t b X$ observable, meaning that in this border ${\cal S}(t\bar t b X)=1$.  On the down left side there is a constraint by direct $t\bar t$ resonance searches: ${\cal S}(t\bar t)> {\cal S}(t\bar t b X)$.   Notice that Fig.~\ref{main} is just pictorial, and in many cases in the forthcoming figures some of these limits are not present.

Finally, it is important to observe that the shape of the $t\bar t bX$ most sensitive region is the outcome of a non-trivial process that includes the different channel cross-sections, the sensitivity of each one of these channels, and the relative magnitudes of their ratio, for each point in parameter space.  Therefore its qualitative understanding could be not simple.

\begin{figure}[ht!]
\begin{minipage}[c]{0.50\textwidth}
\includegraphics[width=\textwidth]{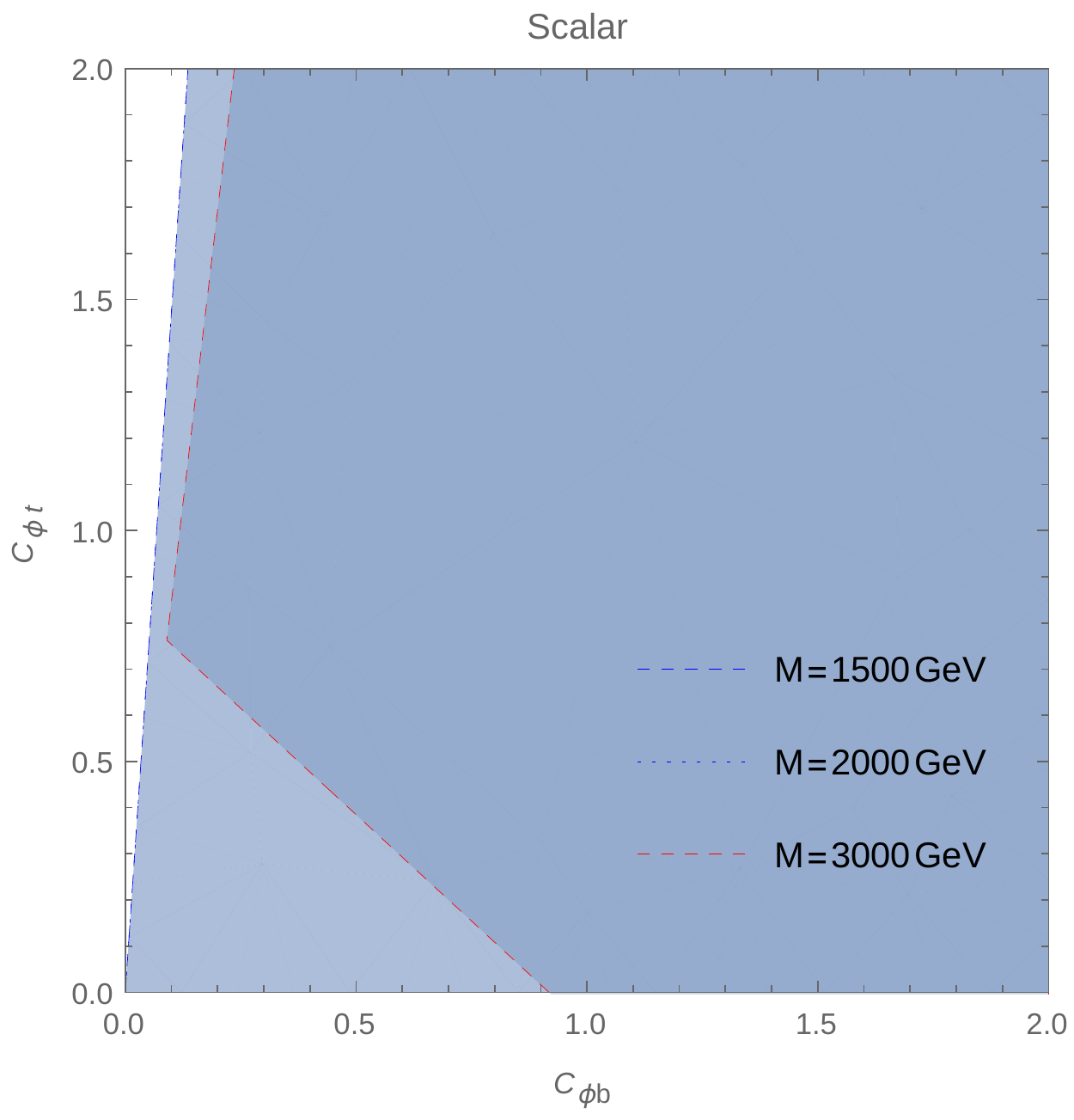}
\end{minipage}
\begin{minipage}[c]{0.50\textwidth}
\includegraphics[width=\textwidth]{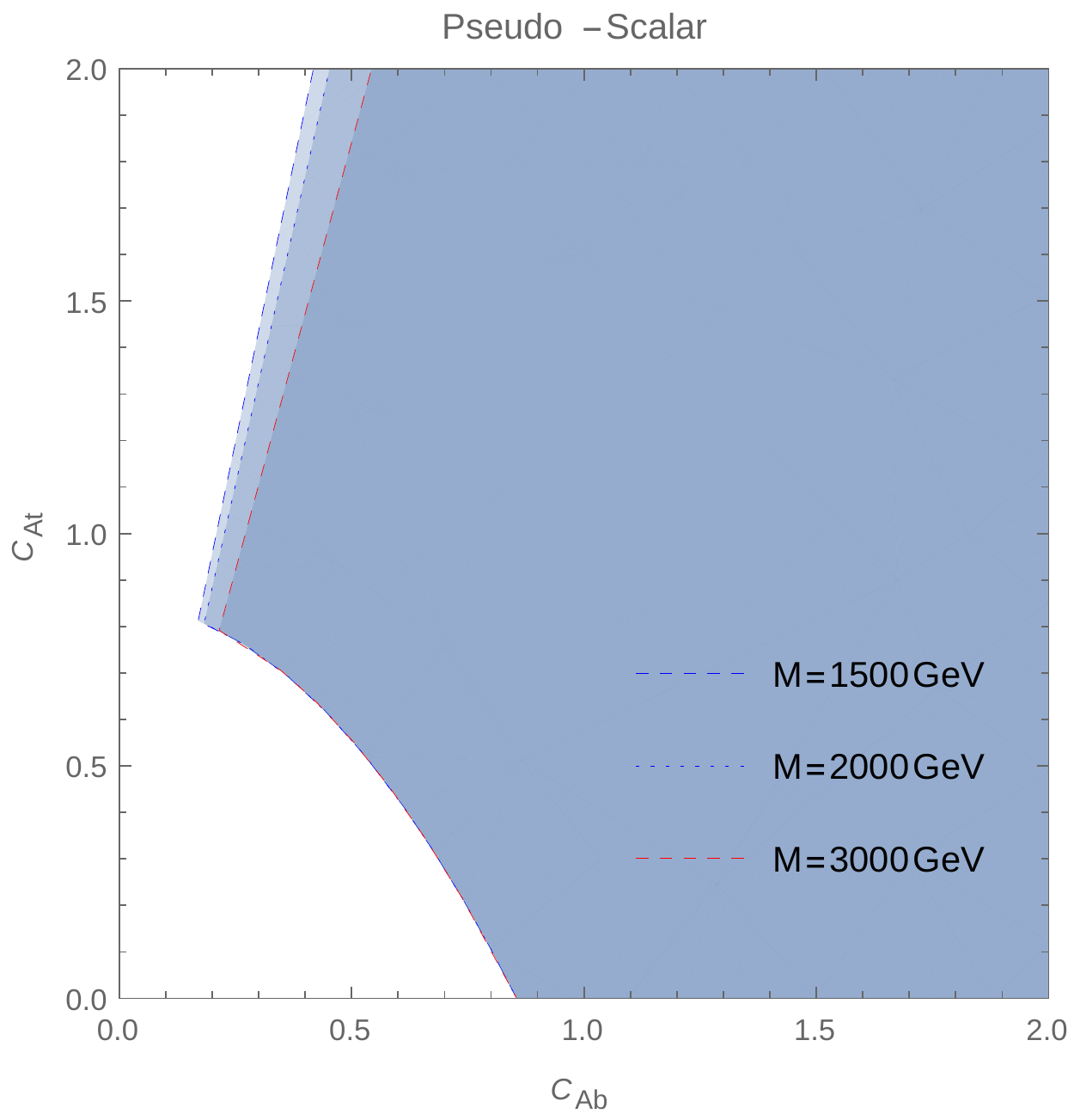}
\end{minipage}
\\
\begin{minipage}[c]{0.50\textwidth}
\includegraphics[width=\textwidth]{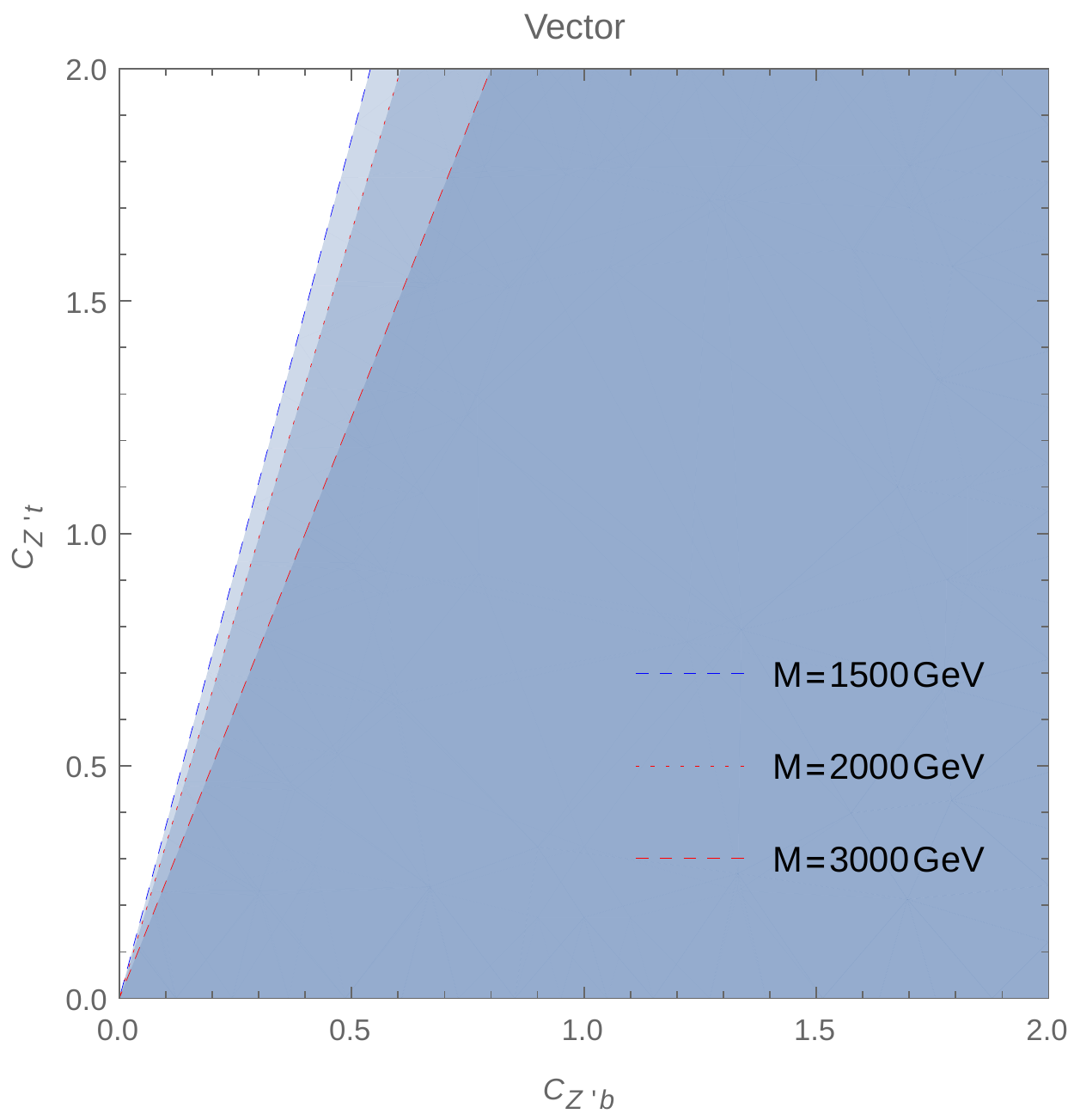}
\end{minipage}
\begin{minipage}[c]{0.50\textwidth}
\includegraphics[width=\textwidth]{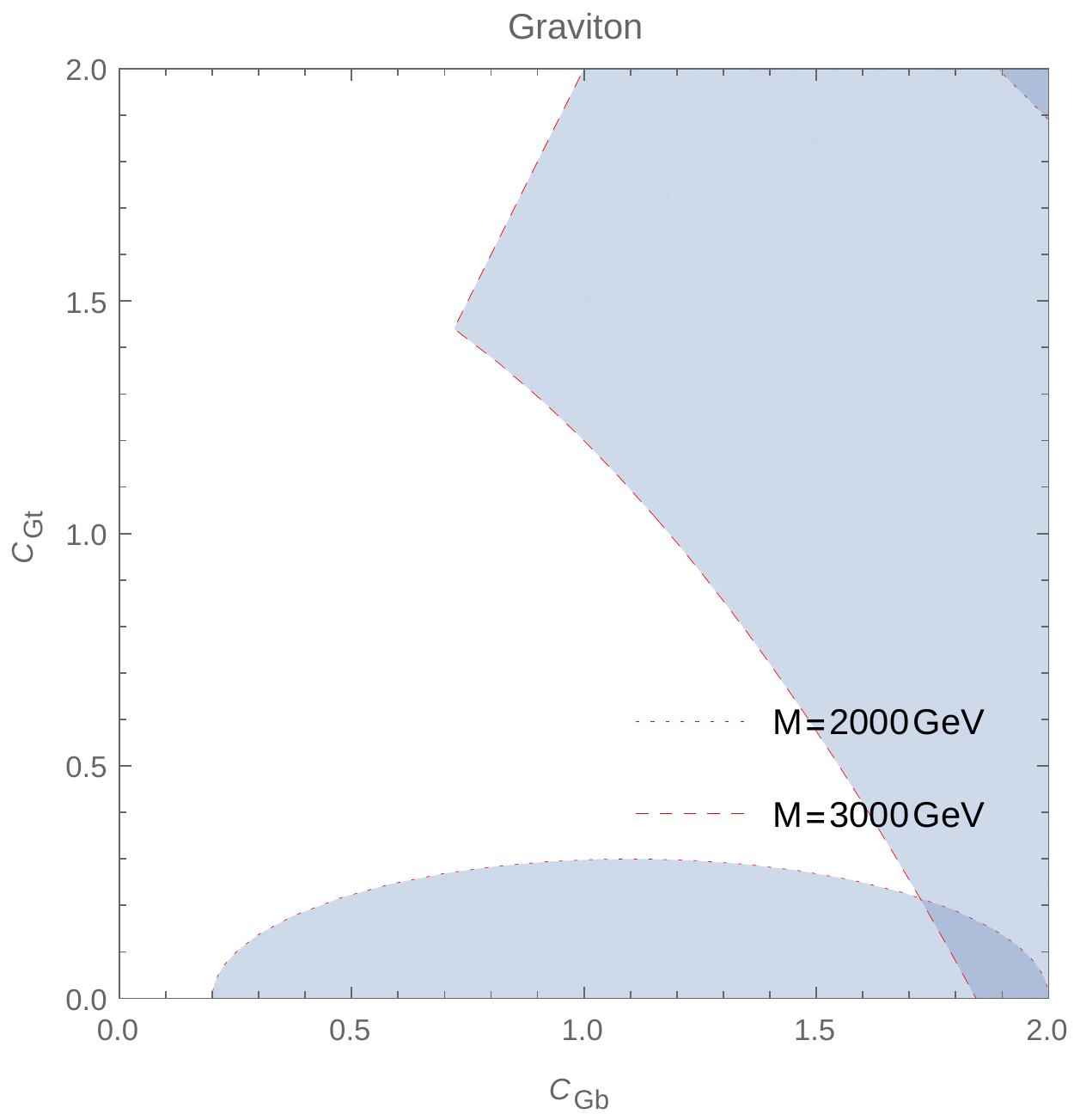}
\end{minipage}
\caption{ \small $t\bar t b X$ most sensitive region for large masses $ M > 1$ TeV.  See text for details.
}
\label{Mlarge}
\end{figure}

We plot in Fig.~\ref{Msmall} the region in parameter space in which $t\bar t b X$ is the most sensitive channel for the cases $M<2m_t$.  As discussed in the previous section, the $b\bar b b X$ channel covers most of the parameter space because of its enhancement due to the inclusiveness that is not present in $t\bar t b X$ because $M<2m_t$ and $Y$ must decay to $b\bar b$.  We also notice an important coverage of the $t\bar t t \bar t$ in the graviton case, as its coupling to fermions includes a derivative term that gets enhanced with the top mass, and suppressed with the bottom mass when fermions are on shell.

In Fig.~\ref{Mmedium} we plot the intermediate mass regions, $2m_t < M < 1$ TeV.  Contrary to the previous case, now the $t\bar t b X$ process is enhanced because the missing $b$ may come from the gluon splitting of diagram D3 in Fig.~\ref{diagrams} and the resonance decay to $t\bar t$.  Therefore, $t\bar t b X$ gains region against the $b\bar b b X$ final state as it is easily seen in the figure.  Another important feature is that the spin-1 $Z'$ is considerably more discarded than the spin-0 and spin-2 cases.  This should be because, as discussed in previous section, the final state with a spin-1 resonance can match, for orbital angular momentum $\ell=0$, both initial states of two gluons with total angular momentum 0 and 2, and therefore is easier to explore/discard.  Finally, we notice again that the $t\bar t t \bar t$ final state has larger impact on the graviton NP model.

The results for the large mass region $M > 1$ TeV are found in Fig.~\ref{Mlarge}. The most important feature of this region is that the $b\bar b b X$ observable has no results due to the difficulty of tagging hard bottoms.  This enlarges the $t\bar t bX$ most sensitive region all the way through the bottom right side, although of course the strength is expected to be quite small in regions where the top coupling is very small.  As a special feature in this mass region, we found that the graviton case has its most important constraint coming from $t\bar t$ resonance searches.  In fact, for $M=1.5$ TeV the whole parameter space in the figure is covered by $t\bar t$ searches (this cannot be seen from the figure), whereas for $2$ and $3$ TeV the sensitivity in $t\bar t$ is stagnant (see \cite{Aad:2015fna}) and $t\bar t b X$ gains relevance.

\section{Discussion}
\label{sec:discussion}

Along the previous section we have shown that with the available experimental searches the final state $t\bar t b X$ is expected to be more sensitive than all other relevant final states in a considerable region in parameter space for the NP lagrangians proposed in Sect.~\ref{sec:model}.  However, many features in the study of the final state $t\bar t b X$ are still in the process of being further understood, and many points should be discussed in this sense.  In this section we address some of this points and also propose some other new items which are relevant for the discussion of this final state.

\subsection{Montecarlo in ttbb}

The first measurements for $t\bar t b \bar b$ at the LHC performed in Refs.~\cite{Aad:2013tua,CMS:2014yxa} had both an excess of about twice the predicted cross-section, with a significance ranging approximately $\sim 1-2\ \sigma$.  In this excess the Montecarlo simulation of the SM process, its hadronization, and the expected cross-section are a crucial input.  In particular, it is well known that the treatment of the gluon splitting $g\to b\bar b$ is decisive for the 4 versus 5 flavor scheme \cite{Maltoni:2012pa}, the factorization scale and the collinear divergences, yielding an important source of uncertainty and also an eventual bias.

In the revision of the above cited references both collaborations ATLAS and CMS have taken different paths in addressing this issue with the Montecarlo prediction.  The ATLAS collaboration has published in Ref.~\cite{Aad:2015yja} a work with special focus on the Montecarlo simulations, in particular they discuss on how to tune Pythia variables that model the $g\to b \bar b$ splitting.  Yielding a better agreement than in their previous work in Ref.~\cite{Aad:2013tua}.  However, one should be cautious, because also NP effects could be hidden in disagreements in the $b\bar b$ distribution of events.  On the other hand, CMS collaboration published in Ref.~\cite{Khachatryan:2015mva} a continuation of Ref.~\cite{CMS:2014yxa}, but now including differential measurements in $b\bar b$ which could contribute to distinguish NP effects observable such as $m_{b\bar b}$ and $\Delta R_{b\bar b}$ among others.

At this level it is suitable to distinguish observables such as the cross-section, whose prediction has a strong dependence in Montecarlo simulations, to other differential observables such as $m_{b\bar b}$ or $\Delta R_{b\bar b}$ in which the kinematic itself can predict a bump in the presence of NP.  These bumps can in principle be distinguished by using the data in the side bands (in $m_{b\bar b}$ or $\Delta R_{b\bar b}$ distributions for instance) and, therefore, we consider these observables to have less Montecarlo impact when compared to others as for instance a fiducial cross-section.  These observables are more reliable at the price of requiring more statistics.

In the previous discussion, the CMS approach seems more solid in their conclusions, however the ATLAS path requires less statistics.  In fact, the CMS approach needs in most of the cases to fully reconstruct the $t\bar t b \bar b$ final state, which yields a great reduction in the number of events.  Given the difficulties in generating a predictable distribution using Montecarlo simulations, we consider that a fully reconstruction of the $t\bar t b \bar b$ event and the search for observables with a reduced impact of Montecarlo generators is a better approach to look for NP in this particular final state, even though the lack of statistics would be one of the main issues.

\subsection{Proposed observable}

In addition to the well studied observables $m_{b\bar b}$ and $\Delta R_{b\bar b}$, which would respectively yield a bump or an excess in back to back events in case of a resonance, we propose a different observable which can also be understood using kinematics.

We consider the general case of a resonance decaying to two particles and study the angle between these two particles 3-momentum.  (For simplicity we consider these particles of equal mass ($m$), however it could be adapted for different mass case.)  In the resonance frame of reference the decay products are back to back, but in the lab reference frame, the decay products get closer due to a Lorentz boost in the direction of the new particle momentum.  From simple kinematics one can see that in the lab frame there is a minimum angle between the 3-momentum of these particles as a function of the resonance momentum and mass.  This minimum angle occurs when the resonance momentum is orthogonal to the decay products direction in the resonance reference frame:
\begin{eqnarray}
\Delta \theta_{min} &=& 2 \arctan \left( \frac{M}{|\vec p |} \sqrt{1 - \left(\frac{2m}{M}\right)^2 }  \right) \label{thetamin} \\
&\approx&  2 \arctan \left( \frac{M}{|\vec p |} \right) \qquad \mbox{(massless case).} \label{thetamin2}
\end{eqnarray}
Where $\vec p$ is the resonance momentum in the Lab Frame, or equivalently the added momentum of the 2 decay products.

In the case of the $t\bar t b \bar b$ final state, if one of the quark pairs comes from a new resonance, then is likely that this new resonance has some transverse momentum since it has been produced in association with other particles. In this case, we can adapt the above formulae for variables more suitable for colliders, in particular for the massless case we propose 
\begin{eqnarray}
\Delta R_{min} &\approx&  2 \arctan \left( \frac{M}{p_T} \right) \qquad  \mbox{(massless case).} \label{rmin2}
\end{eqnarray}

\begin{figure}[h!]
\centering
\begin{minipage}[c]{.47\textwidth}
\centering
\includegraphics[width=\textwidth]{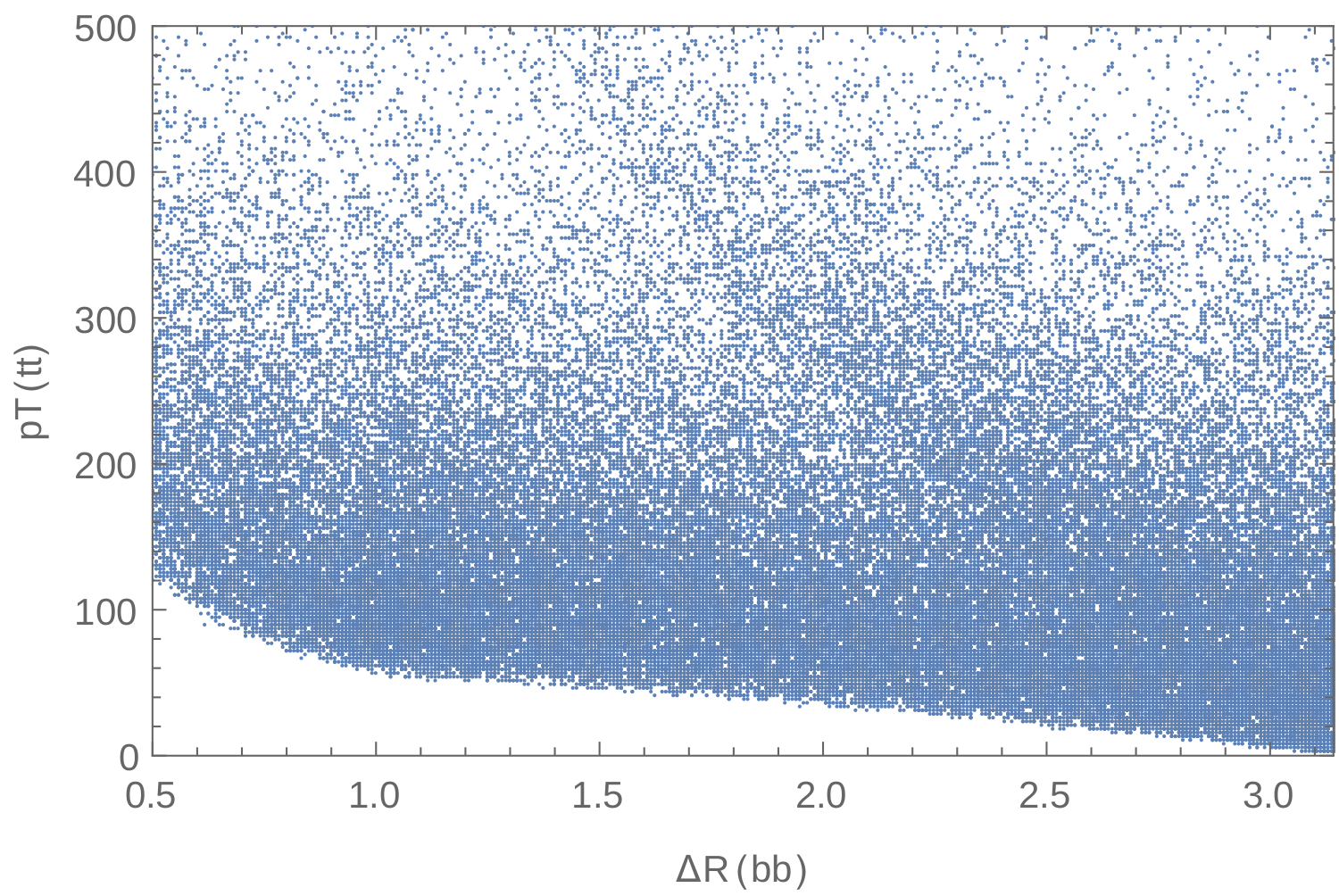} \\
~ ~ (a)
\end{minipage}
\begin{minipage}[c]{.47\textwidth}
\centering
\includegraphics[width=\textwidth]{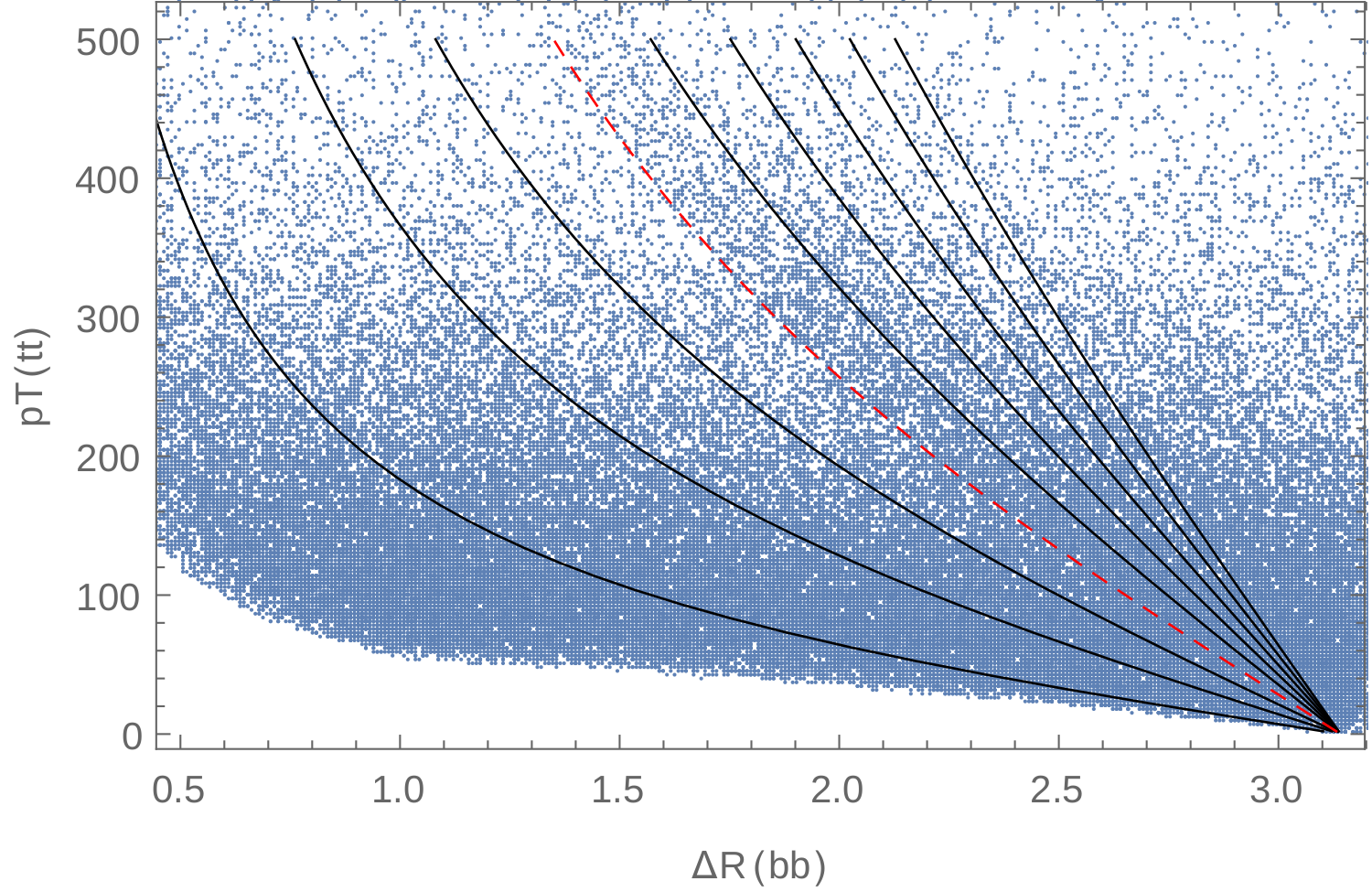} \\
~ ~ (b)
\end{minipage}
\caption{\small (a) Distribution of events in $p p \to t \bar t b \bar b$ for SM+NP.  The NP produces $p p \to t \bar t\, Y,\ Y \to b \bar b$ events which, according to the discussion in text, generates an accumulation of events as it can be seen in the figure.  (b) The same as in the left panel, but with superimposed contour curves of $M_{boosted}$ every $100$ GeV (see Eq.~\ref{mboosted}).  In red-dashed is the contour curve that corresponds to the resonance mass $M=400$ GeV. }
\label{observable}
\end{figure}

\begin{figure}
\centering
\begin{minipage}[c]{.9\textwidth}
\includegraphics[width=\textwidth]{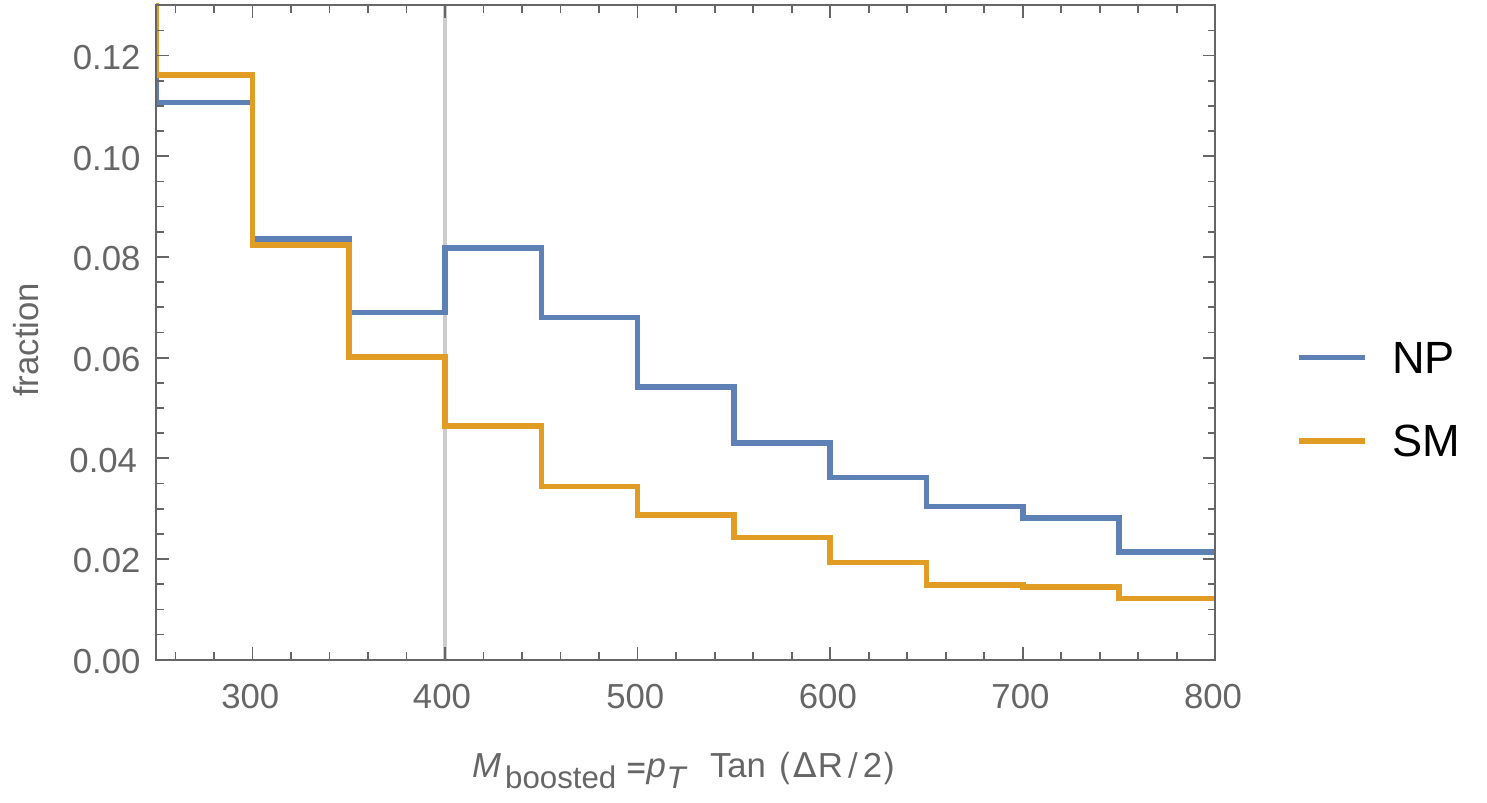}
\end{minipage}
\caption{\small Binning of Fig.~\ref{observable}b.  The correct binning using $M_{boosted}$ in the $p_T(Y)-\Delta R_{b\bar b}$ plane enhances a peak if resonant NP is present in the distribution of events.  The shape of the peak contains information of the angular distribution of the resonance decay products as a function of the resonance momentum in the Lab Frame.  (For better visualization the binning is every $50$ GeV instead of $100$ GeV as in Fig.~\ref{observable}b.)}
\label{histogram}
\end{figure}

Given the above observation, it is interesting to discuss which variables should be measured and compared to enhance the visibility of a possible signal. It is immediate to realize that the new particle decaying to, for instance $Y\to b\bar b$, would yield an accumulation of events in the $p_T(Y)-\Delta R_{b\bar b}$ plane, since the SM background is not expected to have such a kinematic constraint as in Eq.~\ref{rmin2}.  We show in Fig.~\ref{observable}a the distribution of events in the given plane for a case of SM+NP production in $pp\to t\bar t b \bar b$.  Since the accumulation of events follows the curve predicted in Eq.~\ref{rmin2}, it is suitable to bin the events in the plane $p_T(Y)-\Delta R_{b\bar b}$ accordingly for different resonance mass.  By defining the new variable
\begin{equation}
M_{boosted}= p_T \tan \left(\frac{ \Delta R }{2} \right)
\label{mboosted}
\end{equation}
we can conveniently visualize the relevant information of Fig.~\ref{observable}a as in Fig.~\ref{observable}b, and then create a one-dimensional binned plot where the resonance can be easily distinguished, as in Fig.~\ref{histogram}.  In this figure the events in between the curves in Fig.~\ref{observable}b have been collected in definite bins in $M_{boosted}$.  A bump in this new variable, would be a NP signal with reduced impact of Montecarlo simulations; similarly to what happens with the $m_{b\bar b}$ and $\Delta R_{b\bar b}$ observables.

Some considerations on $M_{boosted}$ should be discussed at this point.  The first thing to notice is that, as it can be easily seen, $M_{boosted}$ is neither not equivalent to the invariant mass $m_{b\bar b}$ nor the transverse mass $m_T$ \cite{aaa} of the decay products.  In fact, to reconstruct $M_{boosted}$ are required the decay products direction and transverse momentum, which is not enough to reconstruct the invariant mass.    Analogously, it is not the transverse mass $m_T$, since $M_{boosted}$ requires the longitudinal angular separation $\Delta \eta$ of the decay products and does not have an upper end-point.  It should also be noticed that the $M_{boosted}$-distribution has information on the angle between the decay product directions in the CM frame and the resonance momentum in the lab frame.  It can be shown that scenarios where this angle tends to be orthogonal have a sharper peak in the $M_{boosted}$-distribution.  On the contrary, if this angle tends to be 0 or $\pi$, then the peak in the $M_{boosted}$-distribution is spread.

It would be interesting to investigate whether $M_{boosted}$ could have experimental advantages over other observables as for instance $m_{b\bar b}$ and/or $\Delta R_{b\bar b}$.  In fact, since it requires only the direction and transverse momentum of the decay products, it could have less uncertainty than $m_{b\bar b}$.  Whereas from Fig.~\ref{observable}b one can understand $M_{boosted}$ as the extension of $\Delta R_{b\bar b}$ to the plane for the case of associate resonance production.  This, and other experimental aspects of $M_{boosted}$, are beyond the scope of this work and should be addressed in a separate study.


\subsection{Reach estimate using preliminary 13 TeV results}

As a final point to discuss, we include preliminary results in $t\bar t t \bar t$ and $t\bar t b \bar b$ at 13 TeV, and do a raw estimation of the reach of these observables in the insofar explored NP parameter space.  

In Ref.~\cite{ATLAS:2016btu} a 95\% C.L.~limit is presented in the production cross-section times branching ratio of $pp \to t\bar t Y,\ Y \to t \bar t$ and $pp \to b\bar b Y,\ Y \to t \bar t$ for the intermediate mass region.  These results have still large statistical uncertainties due to the little number of reconstructed events.  We can therefore approximate the scaling of the strength with the luminosity as ${\cal S} \propto \sqrt{L}$ \cite{Alvarez:2016ljl} and obtain a raw estimate of the reach at larger luminosities by finding the ${\cal S } =1$ contour levels after the scaling.

\begin{figure}[h!]
\centering
\begin{minipage}[c]{.47\textwidth}
\centering
\includegraphics[width=\textwidth]{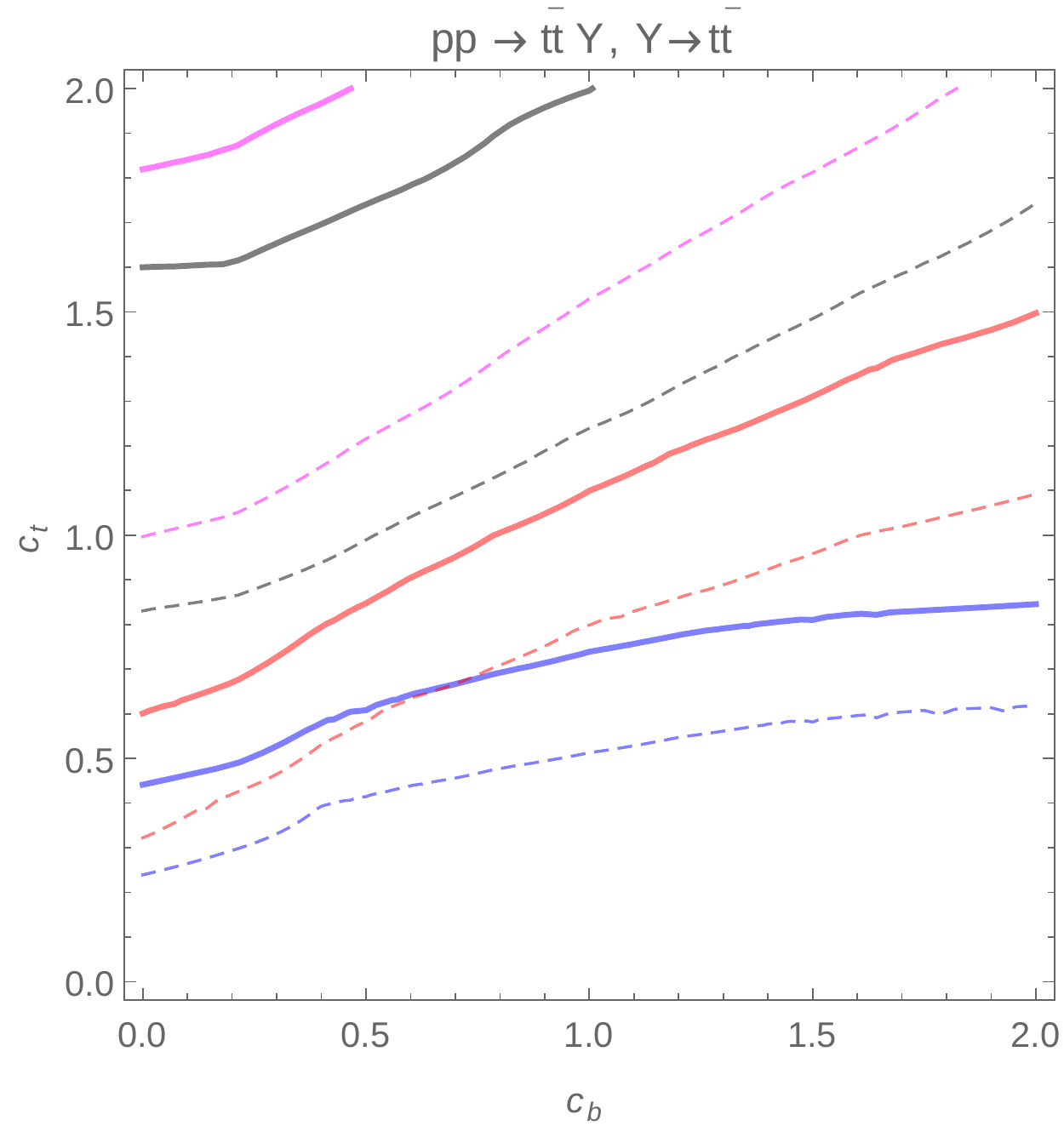} \\
~ ~ ~(a)
\end{minipage}
\begin{minipage}[c]{.47\textwidth}
\centering
\includegraphics[width=\textwidth]{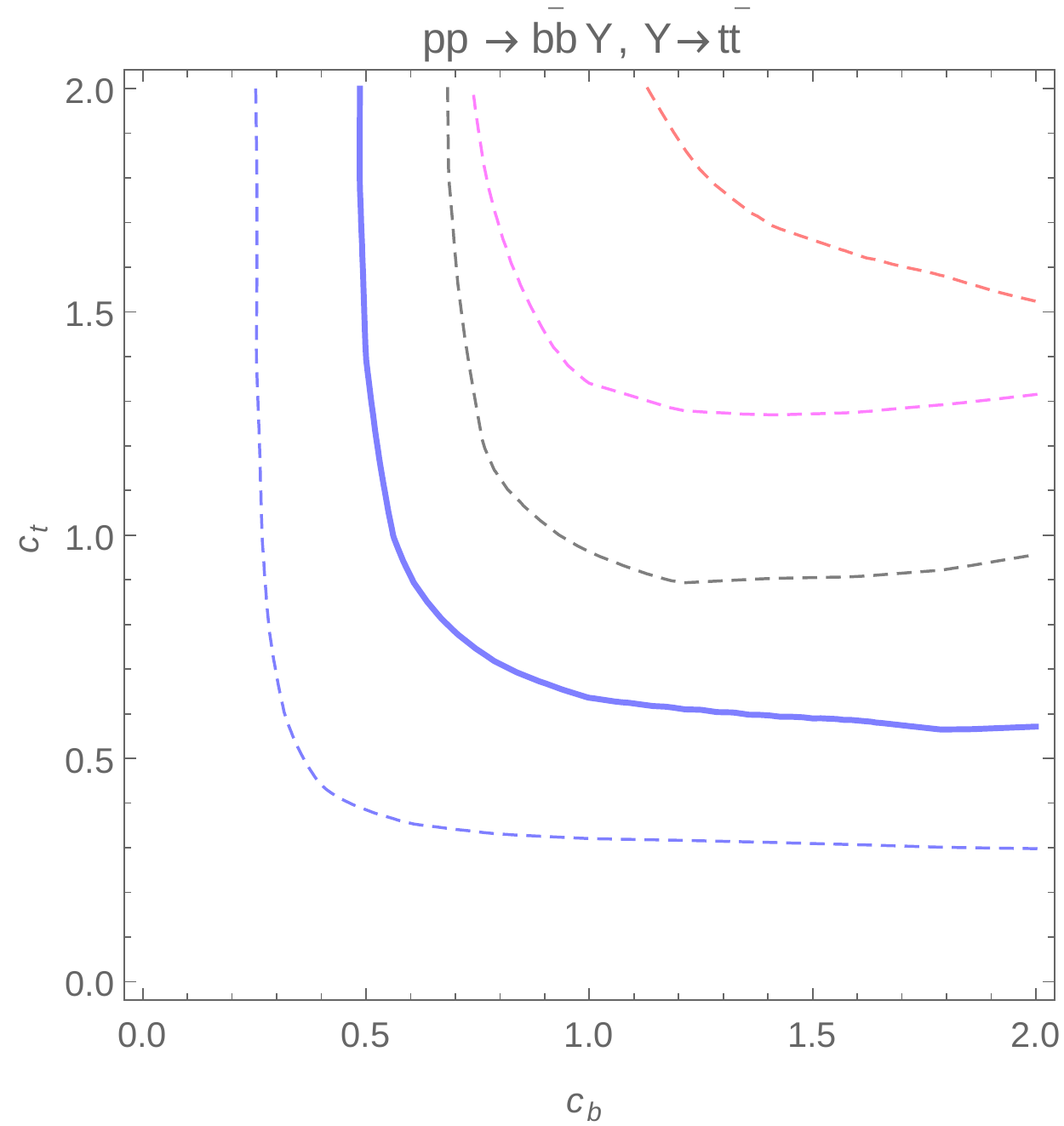} \\
~ ~ ~(b)
\end{minipage}
\caption{\small (a) Reach estimate for the limits in $pp \to t\bar t Y,\ Y \to t \bar t$ in the different NP models of spin 0 scalar (magenta), pseudoscalar (black), vector (blue) and graviton (red), for $M=500$ GeV.  Solid and dashed lines corresponds to 300 fb$^{-1}$ and 3000 fb$^{-1}$, respectively. (b) Similar to left panel, but for the process $pp \to b\bar b Y,\ Y \to t \bar t$.  These limits come from a preliminary result and are expected to improve in many aspects, as discussed in text.}
\label{reach}
\end{figure}

We present in Fig.~\ref{reach} the estimated reach for the 13 TeV observables for the $M=500$ GeV case and for the four different NP models presented in Sect.~\ref{sec:model}.  As expected, the  $pp \to t\bar t Y,\ Y \to t \bar t$ limit has a better constraint for large top couplings and the  $pp \to b\bar b Y,\ Y \to t \bar t$ observable for large top and bottom couplings.  Notice that, as discussed in Sect.~\ref{sec:model}, the $Z'$ model is the easiest to explore/discard due to its spin.  Also notice in the figure how the graviton derivative coupling produces an interplay with the final state fermion masses: there is more sensitivity to tops, and less to bottoms.

The observables as proposed in Ref.~\cite{ATLAS:2016btu} have a reduced impact of Montecarlo generators, requiring in this way more statistics.  This is why their reach is not as restrictive as one would expect due to the energy upgrade.  

It is worth pointing out that according to our discussion in Sect.~\ref{sec:model}, confirmed in the results in Sect.~\ref{sec:ttbb}, the limit in $pp \to b\bar b Y,\ Y \to t \bar t$ would be considerably improved if the experimental analysis would not require to identify and tag both bottom quarks.  That is, if the limit would be imposed in $pp \to b X Y,\ Y \to t \bar t$.  Since in this way there would be an enhancement in the signal due to the integration of a $b$ jet beyond the $|\eta|=2.5$ experimental limit.  We consider that in doing this the limit in this observable would have an important improvement.

\section{Conclusions}
\label{sec:conclusions}

Along this article we have tested simplified NP models that couple exclusively to top and bottom quarks in the context of the $pp \to t\bar t b \bar b$ process.  We have explored spin 0, 1 and 2 NP models and find the region in parameter space in which searches in $t\bar t b X$ are expected to be more sensitive that searches in $t\bar t t \bar t$, $b\bar b b X$ and $t\bar t$, as well as the already discarded parameter space. 

We have made a qualitative analysis of the most relevant Feynman diagrams in the NP models as a function of the different features of the models.  We have made the quantitative comparison by computing the corresponding cross-sections and comparing them to the experimental sensitivity in available searches.  We found that many of the results could be understood from the qualitative analysis.

We have discussed potential issues in detecting NP in the $t\bar t b X$ final state through total cross-sections, since Montecarlo generators may fail in reproducing the bottom quark distributions in the final state.  We have proposed a new observable called $M_{boosted}$ which is based in the angular separation of the decay products of a resonance as a function of its momentum. $M_{boosted}$ could be understood as an extension of $\Delta R_{b\bar b}$ for the case of associated production.  We have shown that a resonance will have a special accumulation of events in the $\Delta R-p_T$ plane, and we have shown how this excess can be binned and measured. We show that $M_{boosted}$ is an observable with reduced impact of Montecarlo, since an excess can be understood from the kinematics; just as it happens with observables as the invariant mass or the angle of separation of decay products.  This observable may be useful in a resonance search through this final state, as well as in other contexts.  Further experimental analysis is required to study the usefulness of this observable.

Using preliminary experimental results we have done a raw estimation of the discovery reach for LHC during the next years.  We have proposed simple improvements to the observable which we show along the work that could extend the reach.

This work shows that the study of the $t\bar t b X$ final state is important in constraining NP that couples to third generation of quarks. The most important result is that we have identified an important fraction of parameter space in these NP models that has not been probed yet and in which $t\bar t b X$ is expected to be the most sensitive channel.  The interest in this region of parameter space is further motivated by slight excess found in many related observables.


\section*{Acknowledgments}

We thank Fabio Maltoni, Jernej Kamenik, Ben Kilminster, and Michele Pinamonti for useful discussions on the contents of this work.  This work is supported by ANPCyT PICT-2266-2013 Funding Project.

\section*{Appendix}\label{sec:appendix}
\renewcommand{\theequation}{{\rm{A}}.\arabic{equation}}
\setcounter{equation}{0}

We provide some more details on the NP models presented in Sect.~\ref{sec:model}

\noindent {\bf \large Scalar $\phi$}

\begin{equation}
F(z)= \left\{ \begin{array}{lcc}
\frac{3}{2} z \left( 1+(1-z)\left(-\frac{1}{4} \left[ \log \left( \frac{1+\sqrt{1-z}}{1-\sqrt{1-z}}\right) -i\pi \right] ^2 \right) \right) &   z<1  \\
             \\ \frac{3}{2} z (1+(1-z) \quad \arctan ^2 \left[ \frac{1}{\sqrt{z-1}} \right] ) &  z>1 \\
             \end{array}
   \right.
\end{equation}

\begin{eqnarray}
\Gamma (\phi \rightarrow q \overline{q})&=&c_{\phi q}^2 \frac{3 M}{8 \pi}\beta ^3 (m_q, M),\\
\Gamma (\phi \rightarrow g g) &=&  \frac{\alpha_s ^2 M^3}{144 \pi ^3}
    \left| \frac{c_{\phi t}}{m_t} F ( z_t ) + \frac{c_{\phi b}}{m_b} F ( z_b ) \right| ^2.\\
\beta(m_q, M) &=& \sqrt{1-\frac{4m_q^2}{M^2}}
\end{eqnarray}

\noindent {\bf \large Pseudoscalar $A$}

\begin{equation}
H(z)= z \left\{ \begin{array}{lcc}
 -\frac{1}{4} \left[ \log \left( \frac{1+\sqrt{1-z}}{1-\sqrt{1-z}}\right) -i\pi \right] ^2  &   z<1  \\
             \\ \arcsin ^2 \left( \sqrt{\frac{1}{z}} \right) &  z>1 \\
             \end{array}
   \right.
\end{equation}

\begin{eqnarray}
\Gamma (A \rightarrow q \overline{q})&=& c_{A q}^2 \frac{3 M}{8 \pi}\beta (m_q, M),\\
\Gamma (A \rightarrow g g) &=&  \frac{\alpha_s ^2 M^3}{64 \pi ^3 }
    \left| \frac{c_{A t}}{m_t} H ( z_t ) +  \frac{c_{A b}}{m_b} H ( z_b ) \right| ^2.
\end{eqnarray}

\noindent {\bf \large Vector $Z'$}

\begin{equation}
\Gamma (Z' \rightarrow q_R \bar q_R) = \frac{3 M}{24 \pi} c_{Z' q}^2.
\end{equation}

\noindent {\bf \large Graviton $G$}

\begin{eqnarray}
A_G(z,\mu_0) &=&  -\f{1}{12}\bigg[
-\f{9}{4}z(z+2) [2 \tanh^{-1}(\sqrt{1-z}) - i\pi]^2
\\
&+& 3(5z+4)\sqrt{1-z}[2 \tanh^{-1}(\sqrt{1-z}) - i\pi]
- 39z - 35 -12 \ln\f{\mu_0^2}{m_q^2} \bigg] \,\quad(z<1) \nonumber
\end{eqnarray}

\begin{eqnarray}
A_G(z,\mu_0) &=& -\f{1}{12}\bigg[
\f{9}{4}z(z+2) [2 \tan^{-1}(\sqrt{z-1}) - \pi]^2
\\
&-& 3(5z+4)\sqrt{z-1}[2 \tan^{-1}(\sqrt{z-1}) - \pi]
- 39z  - 35 -12 \ln\f{\mu_0^2}{m_q^2} \bigg]\,\quad(z>1) \nonumber
\end{eqnarray}

Where following Ref.~\cite{Alvarez:2016ljl} we take $\mu_0=M$ which represents the energy scale of the process.

\begin{eqnarray}
\Gamma(G \to q_R\bar q_R) &=& \frac{3 M^3}{320 \pi \Lambda^2} \beta (m_q, M)^{3/2} \left( c_{Gq}^2 \left( 1-\frac{2 m_q}{3M} \right) \right) , \\
\Gamma(G \to gg) &=& \frac{ M^3}{10 \pi \Lambda^2} \frac{\alpha_s^2}{144\pi^2} \left| c_{G t} A_G( z_t, \mu_0 ) +   c_{G b} A_G( z_b, \mu_0 )  \right| ^2.
\end{eqnarray}

\bibliography{biblio}

\providecommand{\href}[2]{#2}\begingroup\raggedright\begin{thebibliography}{10}

\bibitem{Wess:1974tw}
J.~Wess and B.~Zumino, {\it {Supergauge Transformations in Four-Dimensions}},
  {\em Nucl. Phys.} {\bf B70} (1974) 39--50.

\bibitem{Fayet:1976cr}
P.~Fayet and S.~Ferrara, {\it {Supersymmetry}},  {\em Phys. Rept.} {\bf 32}
  (1977) 249--334.

\bibitem{Randall:1999vf}
L.~Randall and R.~Sundrum, {\it {An Alternative to compactification}},  {\em
  Phys. Rev. Lett.} {\bf 83} (1999) 4690--4693,
  [\href{http://xxx.lanl.gov/abs/hep-th/9906064}{{\tt hep-th/9906064}}].

\bibitem{Gunion:1989we}
J.~F. Gunion, H.~E. Haber, G.~L. Kane, and S.~Dawson, {\it {The Higgs Hunter's
  Guide}},  {\em Front. Phys.} {\bf 80} (2000) 1--404.

\bibitem{Frederix:2007gi}
R.~Frederix and F.~Maltoni, {\it {Top pair invariant mass distribution: A
  Window on new physics}},  {\em JHEP} {\bf 01} (2009) 047,
  [\href{http://xxx.lanl.gov/abs/0712.2355}{{\tt arXiv:0712.2355}}].

\bibitem{Langacker:2008yv}
P.~Langacker, {\it {The Physics of Heavy $Z^\prime$ Gauge Bosons}},  {\em Rev.
  Mod. Phys.} {\bf 81} (2009) 1199--1228,
  [\href{http://xxx.lanl.gov/abs/0801.1345}{{\tt arXiv:0801.1345}}].

\bibitem{Carli:2016eiu}
{\bf ATLAS, CMS} Collaboration, T.~Carli, {\it {Recent top physics highlights
  at the LHC}},  {\em Nucl. Part. Phys. Proc.} {\bf 273-275} (2016) 29--42.

\bibitem{Llacer:2015ypa}
{\bf ATLAS} Collaboration, M.~Moreno~Llácer, {\it {Top quark physics in the
  ATLAS detector: summary of Run I results}},  {\em PoS} {\bf CORFU2014} (2015)
  082.

\bibitem{Margaroli:2015kxa}
{\bf ATLAS, CDF, CMS, D0} Collaboration, F.~Margaroli, {\it {Top quark physics
  at hadron colliders}},  {\em Nuovo Cim.} {\bf C38} (2015), no.~1 10.

\bibitem{ATLAS:2016gvq}
{\bf ATLAS} Collaboration, T.~A. collaboration, {\it {Search for resonances in
  the mass distribution of jet pairs with one or two jets identified as
  $b$-jets with the ATLAS detector with 2015 and 2016 data}}, .

\bibitem{Khachatryan:2015sja}
{\bf CMS} Collaboration, V.~Khachatryan et~al., {\it {Search for resonances and
  quantum black holes using dijet mass spectra in proton-proton collisions at
  $\sqrt{s} =$ 8 TeV}},  {\em Phys. Rev.} {\bf D91} (2015), no.~5 052009,
  [\href{http://xxx.lanl.gov/abs/1501.04198}{{\tt arXiv:1501.04198}}].

\bibitem{Olive:2016xmw}
{\bf Particle Data Group} Collaboration, C.~Patrignani et~al., {\it {Review of
  Particle Physics}},  {\em Chin. Phys.} {\bf C40} (2016), no.~10 100001
  (pp.~624--647).

\bibitem{Mankel:2015ugl}
{\bf CMS} Collaboration, R.~Mankel, {\it {Search for Higgs bosons beyond the
  Standard Model in b-quark final states at CMS experiment}},  {\em PoS} {\bf
  EPS-HEP2015} (2015) 125.

\bibitem{Chatrchyan:2013qga}
{\bf CMS} Collaboration, S.~Chatrchyan et~al., {\it {Search for a Higgs boson
  decaying into a b-quark pair and produced in association with b quarks in
  proton–proton collisions at 7 TeV}},  {\em Phys. Lett.} {\bf B722} (2013)
  207--232, [\href{http://xxx.lanl.gov/abs/1302.2892}{{\tt arXiv:1302.2892}}].

\bibitem{Khachatryan:2015tra}
{\bf CMS} Collaboration, V.~Khachatryan et~al., {\it {Search for neutral MSSM
  Higgs bosons decaying into a pair of bottom quarks}},  {\em JHEP} {\bf 11}
  (2015) 071, [\href{http://xxx.lanl.gov/abs/1506.08329}{{\tt
  arXiv:1506.08329}}].

\bibitem{test0}
T.~A. collaboration, {\it {Search for production of vector-like top quark pairs
  and of four top quarks in the lepton-plus-jets final state in $pp$ collisions
  at $\sqrt{s}=13$ TeV with the ATLAS detector}},  ATLAS-CONF-2016-013.

\bibitem{Beck:2016hyi}
{\bf CMS} Collaboration, L.~Beck, {\it {The search for standard model
  four-top-quark production at $\sqrt{s}$ = 13 TeV in the single lepton and
  opposite-sign dilepton channels}},  in {\em {9th International Workshop on
  Top Quark Physics (TOP 2016) Olomouc, Czech Republic, September 19-23,
  2016}}, 2016.
\newblock \href{http://xxx.lanl.gov/abs/1611.09607}{{\tt arXiv:1611.09607}}.

\bibitem{CMS:2016wig}
{\bf CMS} Collaboration, C.~Collaboration, {\it {Search for standard model
  production of four top quarks in proton-proton collisions at 13 TeV}},
  CMS-PAS-TOP-16-016.

\bibitem{Alvarez:2016nrz}
E.~Alvarez, D.~A. Faroughy, J.~F. Kamenik, R.~Morales, and A.~Szynkman, {\it
  {Four Tops for LHC}},  {\em Nucl. Phys.} {\bf B915} (2017) 19--43,
  [\href{http://xxx.lanl.gov/abs/1611.05032}{{\tt arXiv:1611.05032}}].

\bibitem{CMS:2014yxa}
{\bf CMS} Collaboration, V.~Khachatryan et~al., {\it {Measurement of the cross
  section ratio $\sigma_\mathrm{t \bar{t} b \bar{b}} / \sigma_\mathrm{t \bar{t}
  jj }$ in pp collisions at $\sqrt{s}$ = 8 TeV}},  {\em Phys. Lett.} {\bf B746}
  (2015) 132--153, [\href{http://xxx.lanl.gov/abs/1411.5621}{{\tt
  arXiv:1411.5621}}].

\bibitem{Aad:2015yja}
{\bf ATLAS} Collaboration, G.~Aad et~al., {\it {Measurements of fiducial
  cross-sections for $t\bar{t}$ production with one or two additional b-jets in
  pp collisions at $\sqrt{s}$ =8 TeV using the ATLAS detector}},  {\em Eur.
  Phys. J.} {\bf C76} (2016), no.~1 11,
  [\href{http://xxx.lanl.gov/abs/1508.06868}{{\tt arXiv:1508.06868}}].

\bibitem{ATLAS:2016gqb}
{\bf ATLAS} Collaboration, T.~A. collaboration, {\it {Search for four-top-quark
  production in final states with one charged lepton and multiple jets using
  3.2 fb$^{-1}$ of proton-proton collisions at $\sqrt{s}$ = 13 TeV with the
  ATLAS detector at the LHC}},  ATLAS-CONF-2016-020.

\bibitem{Aad:2013tua}
{\bf ATLAS} Collaboration, G.~Aad et~al., {\it {Study of heavy-flavor quarks
  produced in association with top-quark at $\sqrt{s}=7$ TeV with the ATLAS
  detector}},  {\em Phys. Rev.} {\bf D89} (2014), no.~7 072012,
  [\href{http://xxx.lanl.gov/abs/1304.6386}{{\tt arXiv:1304.6386}}].

\bibitem{Khachatryan:2015mva}
{\bf CMS} Collaboration, V.~Khachatryan et~al., {\it {Measurement of $\mathrm
  {t}\overline{\mathrm {t}}$ production with additional jet activity, including
  $\mathrm {b}$ quark jets, in the dilepton decay channel using pp collisions
  at $\sqrt{s} =$ 8 TeV}},  {\em Eur. Phys. J.} {\bf C76} (2016), no.~7 379,
  [\href{http://xxx.lanl.gov/abs/1510.03072}{{\tt arXiv:1510.03072}}].

\bibitem{ATLAS:2016btu}
{\bf ATLAS} Collaboration, T.~A. collaboration, {\it {Search for new phenomena
  in $t\bar{t}$ final states with additional heavy-flavour jets in $pp$
  collisions at $\sqrt{s}=13$ TeV with the ATLAS detector}}, .

\bibitem{CMS:2016tlo}
{\bf CMS} Collaboration, C.~Collaboration, {\it {Measurement of the cross
  section ratio ttbar+bbbar/ttbar+jj using dilepton final states in pp
  collisions at 13 TeV}},  CMS-PAS-TOP-16-010.

\bibitem{Gori:2016zto}
S.~Gori, I.-W. Kim, N.~R. Shah, and K.~M. Zurek, {\it {Closing the Wedge:
  Search Strategies for Extended Higgs Sectors with Heavy Flavor Final
  States}},  {\em Phys. Rev.} {\bf D93} (2016), no.~7 075038,
  [\href{http://xxx.lanl.gov/abs/1602.02782}{{\tt arXiv:1602.02782}}].

\bibitem{Dolan:2016qvg}
M.~J. Dolan, M.~Spannowsky, Q.~Wang, and Z.-H. Yu, {\it {Determining the
  quantum numbers of simplified models in $t\bar{t}X$ production at the LHC}},
  {\em Phys. Rev.} {\bf D94} (2016), no.~1 015025,
  [\href{http://xxx.lanl.gov/abs/1606.00019}{{\tt arXiv:1606.00019}}].

\bibitem{Jo:2015zxa}
Y.~K. Jo, S.~Y. Choi, Y.~J. Roh, and T.~J. Kim, {\it {Study of the top-quark
  pair production in association with a bottom-quark pair from fast simulations
  at the LHC}},  {\em J. Korean Phys. Soc.} {\bf 67} (2015) 807--812,
  [\href{http://xxx.lanl.gov/abs/1506.04818}{{\tt arXiv:1506.04818}}].

\bibitem{Bevilacqua:2014qfa}
G.~Bevilacqua and M.~Worek, {\it {On the ratio of $ t\overline{t} b\overline{b}
  $ and $ t\overline{t} jj $ cross sections at the CERN Large Hadron
  Collider}},  {\em JHEP} {\bf 07} (2014) 135,
  [\href{http://xxx.lanl.gov/abs/1403.2046}{{\tt arXiv:1403.2046}}].

\bibitem{Craig:2013hca}
N.~Craig, J.~Galloway, and S.~Thomas, {\it {Searching for Signs of the Second
  Higgs Doublet}},  \href{http://xxx.lanl.gov/abs/1305.2424}{{\tt
  arXiv:1305.2424}}.

\bibitem{Branco:2011iw}
G.~C. Branco, P.~M. Ferreira, L.~Lavoura, M.~N. Rebelo, M.~Sher, and J.~P.
  Silva, {\it {Theory and phenomenology of two-Higgs-doublet models}},  {\em
  Phys. Rept.} {\bf 516} (2012) 1--102,
  [\href{http://xxx.lanl.gov/abs/1106.0034}{{\tt arXiv:1106.0034}}].

\bibitem{Hespel:2016qaf}
B.~Hespel, F.~Maltoni, and E.~Vryonidou, {\it {Signal background interference
  effects in heavy scalar production and decay to a top-anti-top pair}},  {\em
  JHEP} {\bf 10} (2016) 016, [\href{http://xxx.lanl.gov/abs/1606.04149}{{\tt
  arXiv:1606.04149}}].

\bibitem{Aad:2015fna}
{\bf ATLAS} Collaboration, G.~Aad et~al., {\it {A search for $ t\overline{t} $
  resonances using lepton-plus-jets events in proton-proton collisions at
  $\sqrt{s}=8$ TeV with the ATLAS detector}},  {\em JHEP} {\bf 08} (2015) 148,
  [\href{http://xxx.lanl.gov/abs/1505.07018}{{\tt arXiv:1505.07018}}].

\bibitem{HP2}
M.~Est\'evez and E.~\'Alvarez, ``Np in the ttbb final state at the lhc.''
  \url{http://icas.unsam.edu.ar/talks/2016.09.07_NP_in_the_ttbb_final_state_at_the_LHC.Mariel_Estevez.pdf},
  2016.

\bibitem{Alvarez:2016ljl}
E.~Alvarez, L.~Da~Rold, J.~Mazzitelli, and A.~Szynkman, {\it {Graviton
  resonance phenomenology and a pNGB Higgs at the LHC}},
  \href{http://xxx.lanl.gov/abs/1610.08451}{{\tt arXiv:1610.08451}}.

\bibitem{Alwall:2014hca}
J.~Alwall, R.~Frederix, S.~Frixione, V.~Hirschi, F.~Maltoni, O.~Mattelaer,
  H.~S. Shao, T.~Stelzer, P.~Torrielli, and M.~Zaro, {\it {The automated
  computation of tree-level and next-to-leading order differential cross
  sections, and their matching to parton shower simulations}},  {\em JHEP} {\bf
  07} (2014) 079, [\href{http://xxx.lanl.gov/abs/1405.0301}{{\tt
  arXiv:1405.0301}}].

\bibitem{Alwall:2011uj}
J.~Alwall, M.~Herquet, F.~Maltoni, O.~Mattelaer, and T.~Stelzer, {\it {MadGraph
  5 : Going Beyond}},  {\em JHEP} {\bf 06} (2011) 128,
  [\href{http://xxx.lanl.gov/abs/1106.0522}{{\tt arXiv:1106.0522}}].

\bibitem{Khachatryan:2014sca}
{\bf CMS} Collaboration, V.~Khachatryan et~al., {\it {Search for Standard Model
  Production of Four Top Quarks in the Lepton + Jets Channel in pp Collisions
  at $\sqrt{s}$ = 8 TeV}},  {\em JHEP} {\bf 11} (2014) 154,
  [\href{http://xxx.lanl.gov/abs/1409.7339}{{\tt arXiv:1409.7339}}].

\bibitem{Maltoni:2012pa}
F.~Maltoni, G.~Ridolfi, and M.~Ubiali, {\it {b-initiated processes at the LHC:
  a reappraisal}},  {\em JHEP} {\bf 07} (2012) 022,
  [\href{http://xxx.lanl.gov/abs/1203.6393}{{\tt arXiv:1203.6393}}]. [Erratum:
  JHEP04,095(2013)].

\bibitem{aaa}
{\bf Particle Data Group} Collaboration, C.~Patrignani et~al., {\it {Review of
  Particle Physics}},  {\em Chin. Phys.} {\bf C40} (2016), no.~10 100001
  (Ch.~47).

\end{thebibliography}\endgroup

\end{document}